\documentclass[12pt]{article}

\usepackage{graphicx}
\usepackage{float}
\usepackage{cite}
\usepackage{amsfonts}
\usepackage{amssymb}
\usepackage{xcolor}

\def\gtwid{\mathrel{\raise.3ex\hbox{$>$\kern-.75em\lower1ex\hbox{$\sim$}}}}
\def\ltwid{\mathrel{\raise.3ex\hbox{$<$\kern-.75em\lower1ex\hbox{$\sim$}}}}
\def\square{\kern1pt\vbox{\hrule height 1.2pt\hbox{\vrule width 1.2pt\hskip 3pt
   \vbox{\vskip 6pt}\hskip 3pt\vrule width 0.6pt}\hrule height 0.6pt}\kern1pt}

\begin{document}

\begin{titlepage}

\begin{flushright}
UFIFT-QG-21-02 , CCTP-2021-02
\end{flushright}

\vskip 1cm

\begin{center}
{\bf How Inflationary Gravitons Affect Gravitational Radiation}
\end{center}

\vskip .5cm

\begin{center}
L. Tan$^{1\star}$, N. C. Tsamis$^{2\dagger}$ and R. P. Woodard$^{1\ddagger}$
\end{center}

\vskip .5cm

\begin{center}
\it{$^{1}$ Department of Physics, University of Florida,\\
Gainesville, FL 32611, UNITED STATES}
\end{center}

\begin{center}
\it{$^{2}$ Institute of Theoretical Physics \& Computational Physics, \\
Department of Physics, University of Crete, \\
GR-710 03 Heraklion, HELLAS}
\end{center}

\vspace{0.5cm}

\begin{center}
ABSTRACT
\end{center}
We include the single graviton loop contribution to the linearized  
Einstein equation. Explicit results are obtained for one loop corrections
to the propagation of gravitational radiation. Although suppressed by a 
minuscule loop-counting parameter, these corrections are enhanced by the
square of the number of inflationary e-foldings. One consequence is that
perturbation theory breaks down for a very long epoch of primordial 
inflation. Another consequence is that the one loop correction to the 
tensor power spectrum might be observable, in the far future, after the
full development of 21cm cosmology.

\begin{flushleft}
\end{flushleft}

\vskip .5cm

\begin{flushleft}
$^{\star}$ e-mail: ltan@ufl.edu \\
$^{\dagger}$ e-mail: tsamis@physics.uoc.gr \\
$^{\ddagger}$ e-mail: woodard@phys.ufl.edu
\end{flushleft}

\end{titlepage}

\section{Introduction}

Primordial inflation produces a vast ensemble of infrared gravitons 
\cite{Starobinsky:1979ty,Starobinsky:1985ww}. If the tensor power 
spectrum for modes of co-moving wave number $k$ is $\Delta^2_{h}(k)$, 
then the occupation number for each polarization of wave vector $\vec{k}$ 
is,
\begin{equation}
N(\eta,k) = \frac{\pi \Delta^2_{h}(k)}{64 G k^2} \times a^2(\eta) \; ,
\end{equation}
where $a(\eta)$ is the scale factor at conformal time $\eta$ and $G$ is
Newton's constant. The aim of this paper is to study how these gravitons
change the kinematics of other gravitons.

The background geometry of cosmology can be well described using 
spatially flat, conformal coordinates,
\begin{equation}
ds^2 = a^2(\eta) \Bigl[-d\eta^2 + d\vec{x} \!\cdot\! d\vec{x}\Bigr] \qquad
\Longrightarrow \qquad H \equiv \frac{a'}{a^2} \quad , \quad \epsilon \equiv 
-\frac{H'}{a H^2} \; , \label{geometry}
\end{equation}
where $H(\eta)$ is the Hubble parameter and $\epsilon(\eta)$ is the first
slow roll parameter. The special case of de Sitter ($\epsilon = 0$, constant 
$H$ and $a(\eta) = -1/H\eta$) is often considered as a reasonable paradigm 
for inflation, and is attractive both because we possess analytic expressions 
for the graviton propagator \cite{Tsamis:1992xa,Woodard:2004ut} and because 
there is no mixing with whatever matter fields support inflation 
\cite{Iliopoulos:1998wq,Abramo:2001dc}. On this background the 
quantum-corrected, linearized Einstein equation takes the form,
\begin{equation}
\mathcal{D}^{\mu\nu\rho\sigma} h_{\rho\sigma}(x) - \int \!\!
d^4x' \Bigl[\mbox{}^{\mu\nu} \Sigma^{\rho\sigma}\Bigr](x;x') h_{\rho\sigma}(x')
= \frac12 \kappa T^{\mu\nu}_{\rm lin}(x) \; , \label{Einsteineqn}
\end{equation}
where $\mathcal{D}^{\mu\nu\rho\sigma}$ is the gauge-fixed kinetic operator,
$-i [\mbox{}^{\mu\nu} \Sigma^{\rho\sigma}](x;x')$ is the 1PI (one particle 
irreducible) 2-graviton function, $T^{\mu\nu}_{\rm lin}(x)$ is the linearized
stress-energy tensor, $\kappa^2 \equiv 16 \pi G$ is the loop-counting
parameter of quantum gravity and $h_{\mu\nu} \equiv (g_{\mu\nu} - a^2 
\eta_{\mu\nu})/\kappa$ is the graviton field. In this paper we will show how 
to extract a fully renormalized result for the one loop graviton self-energy 
from an old, unregulated computation \cite{Tsamis:1996qk}. We then use this 
result to derive one loop corrections to the mode functions of dynamical 
gravitons. 

In section 2 we derive a renormalized result for the part of the graviton 
self-energy which affects dynamical gravitons. Section 3 solves equation 
(\ref{Einsteineqn}) for one loop corrections to transverse-traceless, 
spatial plane waves. Our conclusions comprise section 4. There we discuss 
the exciting possibility that the one loop correction we find might be 
observable.

\section{Quantum Linearized Einstein Equation}

The purpose of this section is to give an explicit expression for the 
linearized Einstein equation (\ref{Einsteineqn}). We begin by defining the
gauge-fixed kinetic operator $\mathcal{D}^{\mu\nu\rho\sigma}$, explaining
how we represent the graviton self-energy, and working out $3+1$
decompositions of both. We then describe the process through
which an old, unregulated computation of the graviton self-energy
\cite{Tsamis:1996qk} can be used to recover most of the fully renormalized, 
Schwinger-Keldysh result. The section closes with a direct computation of 
the remaining contribution.

\subsection{$3+1$ Decomposition}

In $D = 3+1$ dimensions the gauge-fixed kinetic operator is 
\cite{Tsamis:1992xa,Woodard:2004ut},
\begin{equation}
\mathcal{D}^{\mu\nu\rho\sigma} = \frac12 \eta^{\mu (\rho} \eta^{\sigma )\nu}
D_A - \frac14 \eta^{\mu\nu} \eta^{\rho\sigma} D_A + 2 a^4 H^2 \delta^{(\mu}_{~~0}
\eta^{\nu ) (\rho} \delta^{\sigma)}_{~~0} \; , \label{kinop}
\end{equation}
where the kinetic operator of a massless, minimally coupled scalar is,
\begin{equation}
D_A = -a^2 \Bigl[ \partial_0^2 + 2 a H \partial_0 - \nabla^2\Bigr] \; .
\label{DAdef}
\end{equation}
The $3+1$ components of $\mathcal{D}^{\mu\nu\rho\sigma} h_{\rho\sigma}$ are,
\begin{eqnarray}
\mathcal{D}^{00\rho\sigma} h_{\rho\sigma} & = & \frac14 D_A (h_{00} + h_{kk})
- 2 a^4 H^2 h_{00} \; , \qquad \label{D00} \\
\mathcal{D}^{0i\rho\sigma} h_{\rho\sigma} & = & -\frac12 D_B h_{0i} \; , 
\label{D0i} \\
\mathcal{D}^{ij\rho\sigma} h_{\rho\sigma} & = & \frac12 D_A \Bigl[ h_{ij} +
\frac12 \delta_{ij} (h_{00} - h_{kk})\Bigr] \; , \qquad \label{Dij}
\end{eqnarray}
where $D_B$ is the kinetic operator of a conformally coupled scalar,
\begin{equation}
D_B = -a^2 \Bigl[ \partial_0^2 + 2 a H \partial_0 - \nabla^2 + 2 a^2 H^2\Bigr] 
\; . \label{DBdef}
\end{equation}

On a cosmological background (\ref{geometry}) the graviton self-energy can be
expressed as a sum of 21 tensor differential operators $[\mbox{}^{\mu\nu} 
\mathcal{D}^{\rho\sigma}]$ acting on scalar functions of $\eta$, $\eta'$ and
$\Vert \vec{x} - \vec{x}'\Vert$ \cite{Tan:2021ibs},
\begin{equation}
-i\Bigl[\mbox{}^{\mu\nu} \Sigma^{\rho\sigma}\Bigr](x;x') = \sum_{i=1}^{21}
\Bigl[\mbox{}^{\mu\nu} \mathcal{D}_{i}^{\rho\sigma}\Bigr] \!\times\! T^i(x;x') 
\; . \label{initialrep}
\end{equation}
By general tensor analysis the 21 basis tensors are constructed from 
$\delta^{\mu}_{~0}$, the spatial part of the Minkowski metric 
$\overline{\eta}^{\mu\nu} \equiv \eta^{\mu\nu} + \delta^{\mu}_{~0} 
\delta^{\nu}_{~0}$ and the spatial derivative operator $\overline{\partial}^{\mu} 
\equiv \partial^{\mu} + \delta^{\mu}_{~0} \partial_0$. Table~\ref{Tbasis} lists
the $[\mbox{}^{\mu\nu} \mathcal{D}_i^{\rho\sigma}]$.
\begin{table}[H]
\setlength{\tabcolsep}{8pt}
\def\arraystretch{1.5}
\centering
\begin{tabular}{|@{\hskip 1mm }c@{\hskip 1mm }||c||c|c||c|c|}
\hline
$i$ & $[\mbox{}^{\mu\nu} \mathcal{D}^{\rho\sigma}_i]$ & $i$ & $[\mbox{}^{\mu\nu} 
\mathcal{D}^{\rho\sigma}_i]$ & $i$ & $[\mbox{}^{\mu\nu} \mathcal{D}^{\rho\sigma}_i]$ \\
\hline\hline
1 & $\overline{\eta}^{\mu\nu} \overline{\eta}^{\rho\sigma}$ & 8 & $\overline{\partial}^{\mu} 
\overline{\partial}^{\nu} \overline{\eta}^{\rho\sigma}$ & 15 & $\delta^{(\mu}_{~~0} 
\overline{\partial}^{\nu)} \delta^{\rho}_{~0} \delta^{\sigma}_{~0}$ \\
\hline
2 & $\overline{\eta}^{\mu (\rho} \overline{\eta}^{\sigma) \nu}$ & 9 & $\delta^{(\mu}_{~~0} 
\overline{\eta}^{\nu) (\rho} \delta^{\sigma)}_{~~0}$ & 16 & $\delta^{\mu}_{~0} \delta^{\nu}_{~0} 
\overline{\partial}^{\rho} \overline{\partial}^{\sigma}$ \\
\hline
3 & $\overline{\eta}^{\mu\nu} \delta^{\rho}_{~0} \delta^{\sigma}_{~0}$ & 10 & 
$\delta^{(\mu}_{~~0} \overline{\eta}^{\nu) (\rho} \overline{\partial}^{\sigma)}$ & 17 & 
$\overline{\partial}^{\mu} \overline{\partial}^{\nu} \delta^{\rho}_{~0} \delta^{\sigma}_{~0}$ \\
\hline
4 & $\delta^{\mu}_{~0} \delta^{\nu}_{~0} \overline{\eta}^{\rho\sigma}$ & 11 & 
$\overline{\partial}^{(\mu} \overline{\eta}^{\nu) (\rho} \delta^{\sigma)}_{~~0}$ & 18 
& $\delta^{(\mu}_{~~0} \overline{\partial}^{\nu)} \delta^{(\rho}_{~~0} 
\overline{\partial}^{\sigma)}$ \\
\hline
5 & $\overline{\eta}^{\mu\nu} \delta^{(\rho}_{~~0} \overline{\partial}^{\sigma)}$ & 12 & 
$\overline{\partial}^{(\mu} \overline{\eta}^{\nu)(\rho} \overline{\partial}^{\sigma)}$ & 
19 & $\delta^{(\mu}_{~~0} \overline{\partial}^{\nu)} \overline{\partial}^{\rho} 
\overline{\partial}^{\sigma}$ \\
\hline
6 & $\delta^{(\mu}_{~~0} \overline{\partial}^{\nu)} \overline{\eta}^{\rho\sigma}$ & 13 & 
$\delta^{\mu}_{~0} \delta^{\nu}_{~0} \delta^{\rho}_{~0} \delta^{\sigma}_{~0}$ & 20 & 
$\overline{\partial}^{\mu} \overline{\partial}^{\nu} \delta^{(\rho}_{~~0} 
\overline{\partial}^{\sigma)}$ \\
\hline
7 & $\overline{\eta}^{\mu\nu} \overline{\partial}^{\rho} \overline{\partial}^{\sigma}$ & 
14 & $\delta^{\mu}_{~0} \delta^{\nu}_{~0} \delta^{(\rho}_{~~0} \overline{\partial}^{\sigma)}$ 
& 21 & $\overline{\partial}^{\mu} \overline{\partial}^{\nu} \overline{\partial}^{\rho} 
\overline{\partial}^{\sigma}$ \\
\hline
\end{tabular}
\caption{\footnotesize The 21 basis tensors used in expression (\ref{initialrep}). 
The pairs $(3,4)$, $(5,6)$, $(7,8)$, $(10,11)$, $(14,15)$, $(16,17)$ and $(19,20)$ 
are related by reflection.}
\label{Tbasis}
\end{table}
\noindent Reflection invariance --- $-i [\mbox{}^{\mu\nu} \Sigma^{\rho\sigma}](x;x') 
= -i [\mbox{}^{\rho\sigma} \Sigma^{\mu\nu}](x';x)$ --- relates the 7 pairs of 
$T^i(x;x')$ given in Table~\ref{ReflectionT}.
\begin{table}[H]
\setlength{\tabcolsep}{8pt}
\def\arraystretch{1.5}
\centering
\begin{tabular}{|@{\hskip 1mm }c@{\hskip 1mm }||c||c|c|}
\hline
$i$ & Relation & $i$ & Relation \\
\hline\hline
$3,4$ & $T^4(x;x') = +T^3(x';x)$ & $14,15$ & $T^{15}(x;x') = -T^{14}(x';x)$ \\
\hline
$5,6$ & $T^6(x;x') = -T^5(x';x)$ & $16,17$ & $T^{17}(x;x') = +T^{16}(x';x)$ \\
\hline
$7,8$ & $T^8(x;x') = +T^7(x';x)$ & $19,20$ & $T^{20}(x;x') = -T^{19}(x';x)$ \\
\hline
$10,11$ & 
$T^{11}(x;x') = -T^{10}(x';x)$ & $$ & $$ \\
\hline
\end{tabular}
\caption{\footnotesize Scalar coefficient functions in expression (\ref{initialrep})
which are related by reflection.}
\label{ReflectionT}
\end{table}

The factor $[\mbox{}^{\mu\nu} \Sigma^{\rho\sigma}](x;x') h_{\rho\sigma}(x')$ in 
expression (\ref{Einsteineqn}) has the $3 + 1$ decomposition,
\begin{eqnarray}
\lefteqn{ \Bigl[\mbox{}^{00} \Sigma^{\rho\sigma}\Bigr] h_{\rho\sigma} \longrightarrow
i T^4 h_{kk} + i T^{13} h_{00} + i T^{14} h_{0k , k} + i T^{16} h_{k \ell , k\ell} 
\; , } \label{Sigma00} \\
\lefteqn{ \Bigl[\mbox{}^{0i} \Sigma^{\rho\sigma}\Bigr] h_{\rho\sigma} \longrightarrow
\frac{i}{2} \partial_i \Bigl[T^6 h_{kk} + T^{15} h_{00} + T^{18} h_{0k ,k} + T^{19}
h_{k\ell ,k\ell}\Bigr] } \nonumber \\
& & \hspace{8cm} + \frac{i}{2} T^9 h_{0i} + \frac{i}{2} T^{10} h_{i k , k} \; , 
\qquad \label{Sigma0i} \\
\lefteqn{ \Bigl[\mbox{}^{ij} \Sigma^{\rho\sigma}\Bigr] h_{\rho\sigma} \longrightarrow
i \delta_{ij} \Bigl[ T^1 h_{kk} + T^3 h_{00} + T^5 h_{0k ,k} + T^7 h_{k\ell ,k \ell}
\Bigr] + i T^2 h_{ij} } \nonumber \\
& & \hspace{-0.5cm} + i \partial_{( i} \Bigl[ T^{11} h_{j) 0} \!+\! T^{12}
h_{j ) k, k}\Bigr] + i \partial_i \partial_j \Bigl[ T^8 h_{kk} \!+\! T^{17} h_{00}
\!+\! T^{20} h_{0k ,k} \!+\! T^{21} h_{k\ell ,k\ell}\Bigr] . \qquad \label{Sigmaij}
\end{eqnarray}
In these relations we have sometimes exploited spatial transition invariance to
partially integrate spatial derivatives from the coefficient functions $T^i(x;x')$ 
onto the graviton field.

\subsection{The Quantum Correction}

\begin{figure}[H]
\centering
\includegraphics[width=11cm]{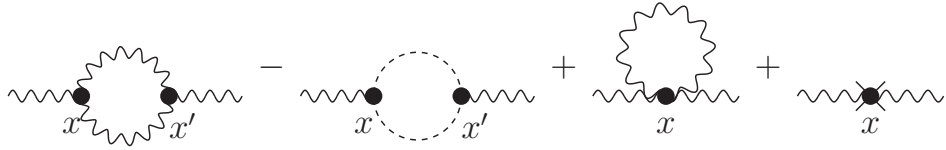}
\caption{\footnotesize Diagrams contributing to the one loop graviton 
self-energy, shown in the same order, left to right, as the four contributions 
to (\ref{operatorexpr}). Graviton lines are wavy and ghost lines are dashed.}
\label{diagrams}
\end{figure}

Let $S[g]$ represent the classical action of gravity, while $S_g[h,\overline{\theta},
\theta]$ stands for the ghost and gauge fixing action, and $\Delta S[g]$ is the 
counter-action. The one loop graviton self-energy can be expressed as the 
expectation value of the sum of four variational derivatives of these quantities,
\begin{eqnarray}
\lefteqn{ -i \Bigl[\mbox{}^{\mu\nu} \Sigma^{\rho\sigma}\Bigr](x;x') = 
\Biggl\langle \Omega \Biggl\vert T^*\Biggl[ \Bigl[\frac{i \delta S[g]}{\delta 
h_{\mu\nu}(x)} \Bigr]_{h h} \Bigl[ \frac{i \delta S[g]}{\delta h_{\rho\sigma}(x')}
\Bigr]_{h h} + \Bigl[\frac{i \delta S[g]}{\delta 
h_{\mu\nu}(x)} \Bigr]_{\overline{\theta} \theta} } \nonumber \\
& & \hspace{0.3cm} \times \Bigl[ \frac{i \delta S[g]}{\delta h_{\rho\sigma}(x')}
\Bigr]_{\overline{\theta} \theta} + \Bigl[ \frac{i \delta^2 S[g]}{\delta 
h_{\mu\nu}(x) \delta h_{\rho\sigma}(x')} \Bigr]_{hh} + \Bigl[ \frac{i \delta^2 
\Delta S[g]}{\delta h_{\mu\nu}(x) \delta h_{\rho\sigma}(x')} \Bigr]_{1} \Biggr] 
\Biggr\vert \Omega \Biggr\rangle . \qquad \label{operatorexpr}
\end{eqnarray}
The various subscripts indicate how many graviton fields contribute and the
$T^*$-ordering symbol signifies that any derivatives are taken after time 
ordering. Figure~\ref{diagrams} gives the relevant Feynman diagrams.
\vskip .5cm

\subsubsection{The $D=4$ Result}

To understand the unregulated result \cite{Tsamis:1996qk} it helps to 
consider how a computation of $-i[\mbox{}^{\mu\nu} \Sigma^{\rho\sigma}](x;x')$ 
would look in dimensional regularization. The 3-graviton and 4-graviton vertices 
take the general form \cite{Tsamis:1992xa,Tsamis:1996qm},\footnote{
Ghost-antighost-graviton vertices are similar to (\ref{vertex3pt}).}
\begin{eqnarray}
\kappa a^{D-2} h \partial h \partial h \qquad & , & \qquad \kappa H a^{D-1} 
h h \partial h \; , \label{vertex3pt} \\
\kappa^2 a^{D-2} h h \partial h \partial h \qquad & , & \qquad \kappa^2 H 
a^{D-1} h h h \partial h \; . \qquad \label{vertex4pt}
\end{eqnarray}
Without worrying about indices (which come in a dizzying variety of different
possibilities), contributions to the first two diagrams of Figure~\ref{diagrams} 
have the nonlocal structure,
\begin{equation}
\kappa a^{D-2} \times \partial \partial' i\Delta(x;x') \times \partial 
\partial' i\Delta(x;x') \times \kappa {a'}^{D-2} \; , \label{generic3pt}
\end{equation}
where $i\Delta(x;x')$ is a ghost or graviton propagator, and one derivative
at each vertex could be exchanged for $H$ times the appropriate scale factor.
One should also bear in mind that if the external leg graviton field happens
to be differentiated then minus its derivative is acted on the entire
diagram. In contrast, the third diagram of Figure~\ref{diagrams} is local,
\begin{equation}
\kappa^2 a^{D-2} \times \partial \partial' i\Delta(x;x') \times
i\delta^D(x \!-\! x') \; , \label{generic4pt}
\end{equation}
with the same stipulation about derivatives. The final diagram of 
Figure~\ref{diagrams} is also local,
\begin{equation}
\frac{\kappa^2 a^{D-4}}{D \!-\! 4} \times \partial^2 {\partial'}^2 \times
i\delta^D(x \!-\! x') \; , \label{genericctm}
\end{equation}
where any number of the four derivatives could be exchanged for $H$ times 
a scale factor.

The graviton and ghost propagators were defined by adding to the Lagrangian
a functional whose $D$-dimensional extension is \cite{Tsamis:1992xa,
Woodard:2004ut},
\begin{equation}
\mathcal{L}_{GF} = -\frac{a^{D-2}}{2} \eta^{\mu\nu} F_{\mu} F_{\nu} 
\;\; , \;\; F_{\mu} = \eta^{\rho\sigma} \Bigl( h_{\mu\rho , \sigma} - 
\frac12 h_{\rho\sigma , \mu} + (D\!-\! 2) a H h_{\mu\rho} \delta^{0}_{~\sigma} 
\Bigr) \; . \label{dSgauge}
\end{equation}
The wonderful property of this gauge is that the gauge and graviton propagators 
are sums of constant tensor factors times scalar propagators,
\begin{eqnarray}
i\Bigl[\mbox{}_{\mu} \Delta_{\rho}\Bigr](x;x') & = & \overline{\eta}_{\mu\rho}
\times i\Delta_A(x;x') - \delta^0_{~\mu} \delta^0_{~\nu} \times i\Delta_B(x;x')
\; , \label{ghostprop} \\
i\Bigl[\mbox{}_{\mu\nu} \Delta_{\rho\sigma}\Bigr](x;x') & = & \sum_{I=A,B,C} 
\Bigl[\mbox{}_{\mu\nu} T^I_{\rho\sigma}\Bigr] \times i\Delta_I(x;x') \; .
\label{gravprop} 
\end{eqnarray}
The constant tensor factors $[\mbox{}_{\mu\nu} T^I_{\rho\sigma}]$ are,
\begin{eqnarray} 
\Bigl[\mbox{}_{\mu\nu} T^A_{\rho\sigma}\Bigr] = 2 \overline{\eta}_{\mu (\rho}
\overline{\eta}_{\sigma) \nu} - \frac{2}{D\!-\!3} \overline{\eta}_{\mu\nu}
\overline{\eta}_{\rho\sigma} \quad , \quad \Bigl[\mbox{}_{\mu\nu} T^B_{\rho\sigma}
\Bigr] = -4 \delta^0_{~(\mu} \overline{\eta}_{\nu) (\rho} \delta^0_{~ \sigma)} 
\; , \qquad \\
\Bigl[\mbox{}_{\mu\nu} T^C_{\rho\sigma}\Bigr] = \frac{2 E_{\mu\nu} E_{\rho\sigma}}{
(D\!-\!2) (D\!-\!3)} \quad , \quad E_{\mu\nu} \equiv (D\!-\!3) \delta^0_{~\mu}
\delta^0_{~\nu} + \overline{\eta}_{\mu\nu} \; . \qquad 
\end{eqnarray}
The three scalar propagators are given in terms of the de Sitter length
function $y(x;x') \equiv a a' H^2 \Delta x^2$ and a function $A(y)$,
\begin{eqnarray}
i\Delta_A(x;x') & = & A(y) + k \ln(a a') \qquad k \equiv 
\frac{H^{D-2}}{(4\pi)^{\frac{D}2}} \frac{\Gamma(D\!-\!1)}{\Gamma(\frac{D}2)} \; , 
\label{DeltaA} \\
i\Delta_B(x;x') & = & B(y) \equiv -\frac{[(4 y \!-\! y^2) A'(y) \!+\! (2 \!-\! y) k]}{
2 (D \!-\! 2)} \; , \label{DeltaB} \\
i\Delta_C(x;x') & = & C(y) \equiv \frac12 (2 \!-\! y) B(y) + \frac{k}{D\!-\!3} \; .
\label{DeltaC}
\end{eqnarray}
The first derivative of $A(y)$ is \cite{Onemli:2002hr,Onemli:2004mb},
\begin{eqnarray}
\lefteqn{A'(y) = -\frac{H^{D-2}}{4 (4 \pi)^{\frac{D}2}} \Biggl\{ \Gamma\Bigl(
\frac{D}2\Bigr) \Bigl( \frac{4}{y}\Bigr)^{\frac{D}2} + \Gamma\Bigl( \frac{D}2 
\!+\! 1\Bigr) \Bigl( \frac{4}{y}\Bigr)^{\frac{D}2 - 1} } \nonumber \\
& & \hspace{2cm} + \sum_{n=0}^{\infty} \Biggl[ \frac{\Gamma(n \!+\! 
\frac{D}2 \!+\! 2)}{\Gamma(n \!+\! 3)} \Bigl( \frac{y}{4}\Bigr)^{n-\frac{D}2 + 2} 
- \frac{\Gamma(n \!+\! D)}{\Gamma(n \!+\! \frac{D}2 \!+\! 1)} \Bigl( 
\frac{y}{4}\Bigr)^{n} \Biggr] \Biggr\} . \qquad \label{Aprime}
\end{eqnarray}
Because the $y^n$ and $y^{n-\frac{D}2 - 2}$ terms cancel in $D=4$ dimensions
they can only contribute when multiplied by an ultraviolet divergence.

Divergences only occur from taking a propagator to coincidence, that is, by
setting ${x'}^{\mu} = x^{\mu}$. It follows that we can take the unregulated
($D=4$) limit in the first two diagrams of Figure~\ref{diagrams} so long as
${x'}^{\mu} \neq x^{\mu}$. Of course that same stipulation makes the last
two diagrams of Figure~\ref{diagrams} vanish. Taking $D=4$ also results in a
huge simplification of the ghost and graviton propagators,
\begin{eqnarray}
i\Bigl[\mbox{}_{\mu} \Delta^{D=4}_{\rho}\Bigr](x;x') & = & \frac1{4\pi^2} \Biggl\{
\frac{\eta_{\mu\rho}}{a a' \Delta x^2} - \frac12 H^2 \ln(H^2 \Delta x^2) 
\overline{\eta}_{\mu\rho} \Biggr\} , \qquad \label{ghostD4} \\
i\Bigl[\mbox{}_{\mu\nu} \Delta^{D=4}_{\rho\sigma}\Bigr](x;x') & = & \frac1{4\pi^2}
\Biggl\{ \frac{(2 \eta_{\mu (\rho} \eta_{\sigma) \nu} \!-\! \eta_{\mu\nu}
\eta_{\rho\sigma})}{a a' \Delta x^2} \nonumber \\
& & \hspace{2cm} - H^2 \ln(H^2 \Delta x^2) \Bigl(\overline{\eta}_{\mu (\rho} 
\overline{\eta}_{\sigma) \nu} \!-\! \overline{\eta}_{\mu\nu} 
\overline{\eta}_{\rho\sigma} \Bigr) \Biggr\} . \qquad \label{gravD4}
\end{eqnarray}
\begin{table}[H]
\setlength{\tabcolsep}{8pt}
\def\arraystretch{1.5}
\centering
\begin{tabular}{|@{\hskip 1mm }c@{\hskip 1mm }||c|}
\hline
$i$ & Coefficients $T^i_{L}(x;x')$ in expression (\ref{TNTLdef}) \\
\hline\hline
$1$ & 
$8 a^2 {a'}^2 H^4 \times [ \frac{4 \Delta \eta^2}{\Delta x^6} + \frac{1}{\Delta x^4}] 
+ 4 a^3 {a'}^3 H^6 \times [ \frac{4 \Delta \eta^4}{\Delta x^6} - \frac{\Delta \eta^2
}{\Delta x^4} + \frac{3}{\Delta x^2}]$ \\
\hline
$2$ & 
$-16 a^2 {a'}^2 H^4 \times [ \frac{4 \Delta \eta^2}{\Delta x^6} + \frac{1}{\Delta x^4}]
- 4 a^3 {a'}^3 H^6 \times [ \frac{8 \Delta \eta^4}{\Delta x^6} + \frac{1}{\Delta x^2}]$ \\
\hline
$3$ & $8 a^3 {a'}^3 H^6 \!\times\! [\frac{\Delta \eta^2}{\Delta x^4} \!-\! 
\frac{2}{\Delta x^2}] \!-\! 4 a^3 {a'}^2 H^5 \!\times\! 
[\frac{4 \Delta \eta^3}{\Delta x^6} \!-\! \frac{\Delta \eta}{\Delta x^4}]$ \\
\hline
$5$ &
$16 a^2 {a'}^2 H^4 \times \frac{\Delta \eta}{\Delta x^4} + 4 a^3 {a'}^2 H^5 \times 
[\frac{2 \Delta \eta^2}{\Delta x^4} + \frac{3}{\Delta x^2}]$ \\
\hline
$7$ &
$-8 a^2 {a'}^2 H^4 \times \frac1{\Delta x^2} - 2 a^3 {a'}^3 H^6 \times 
\frac{\Delta \eta^2}{\Delta x^2} - 2 a^3 {a'}^2 H^5 \times \frac{\Delta \eta}{
\Delta x^2}$ \\
\hline
$9$ &
$-96 a a' H^2 \times [\frac{16 \Delta \eta^4}{\Delta x^{10}} + \frac{12 \Delta \eta^2}{
\Delta x^8} + \frac{1}{\Delta x^6}] - 4 a^2 {a'}^2 H^4 \times [\frac{24 \Delta \eta^4}{
\Delta x^8} + \frac{8 \Delta \eta^2}{\Delta x^6} - \frac{1}{\Delta x^4}]$ \\
\hline
$10$ &
$\!\!96 a a' H^2 \!\times\! [\frac{2 \Delta \eta^3}{\Delta x^8} \!+\! 
\frac{\Delta \eta}{\Delta x^6}] \!+\! 12 a^2 {a'}^2 H^4 \!\times\! 
[\frac{4 \Delta \eta^3}{\Delta x^6} \!+\! \frac{\Delta \eta}{\Delta x^4}]
\!+\! a^3 {a'}^3 H^6 \!\times\! [\frac{8 \Delta \eta^3}{\Delta x^4} \!-\!
\frac{4 \Delta \eta}{\Delta x^2}] \!\!$ \\
$$ & $-8 a^2 a' H^3 \!\times\! [\frac{4 \Delta \eta^2}{\Delta x^6} \!+\! 
\frac1{\Delta x^4}] \!-\! 4 a^3 a'^2 H^5 \!\times\! [\frac{2 \Delta \eta^2}{\Delta x^4} 
\!-\! \frac{1}{\Delta x^2}]$ \\
\hline
$12$ &
$-8 a a' H^2 \!\times\! [\frac{4 \Delta \eta^2}{\Delta x^6} \!+\! \frac{1}{\Delta x^4}]
\!-\! 2 a^2 {a'}^2 H^4 \!\times\! [\frac{6 \Delta \eta^2}{\Delta x^4} \!-\! 
\frac{9}{\Delta x^2}] \!+\! 4 a^3 {a'}^3 H^6 \times \frac{\Delta \eta^2}{\Delta x^2}$ \\
\hline
$13$ & 
$-96 a a' H^2 \times [\frac{16 \Delta \eta^4}{\Delta x^{10}} \!+\! \frac{12 \Delta \eta^2}{
\Delta x^8} \!+\! \frac{1}{\Delta x^6}] + 4 a^2 {a'}^2 H^4 \times [\frac{24 \Delta \eta^4}{
\Delta x^8} \!+\! \frac{56 \Delta \eta^2}{\Delta x^6} \!+\! \frac{11}{\Delta x^4}]$ \\
$$ & $+ 8 a^3 {a'}^3 H^6 \times [ \frac{4 \Delta \eta^4}{\Delta x^6} - 
\frac{2 \Delta \eta^2}{\Delta x^4} + \frac{3}{\Delta x^2}]$ \\
\hline
$14$ & 
$192 a a' H^2 \times [\frac{2 \Delta \eta^3}{\Delta x^8} + \frac{\Delta \eta}{\Delta x^6}]
+ 8 a^2 {a'}^2 H^4 \times [\frac{4 \Delta \eta^3}{\Delta x^6} - \frac{3 \Delta \eta}{
\Delta x^4}]$ \\
$$ & $-16 a^2 a' H^3 \times [\frac{4 \Delta \eta^2}{\Delta x^6} + \frac1{\Delta x^4}]
- 16 a^3 {a'}^2 H^5 \times [\frac{\Delta \eta^2}{\Delta x^4} + \frac{1}{\Delta x^2}]$ \\
\hline
$16$ &
$-8 a a' H^2 \!\times\! [\frac{4 \Delta \eta^2}{\Delta x^6} \!+\! \frac{1}{\Delta x^4}]
\!+\! 2 a^2 {a'}^2 H^4 \!\times\! [-\frac{6 \Delta \eta^2}{\Delta x^4} \!+\! 
\frac1{\Delta x^2}] \!-\! 2 a^3 {a'}^3 H^6 \!\times\! \frac{\Delta \eta^2}{\Delta x^2}$ \\
$$ & $+ 16 a^2 a' H^3 \times \frac{\Delta \eta}{\Delta x^4} + 6 a^3 {a'}^2 H^5 \times
\frac{\Delta \eta}{\Delta x^2}$ \\
\hline
$18$ & 
$-24 a a' H^2 \times [\frac{4 \Delta \eta^2}{\Delta x^6} + \frac{1}{\Delta x^4}]
-2 a^2 {a'}^2 H^4 \times [\frac{6 \Delta \eta^2}{\Delta x^4} + \frac{5}{\Delta x^2}]$ \\
\hline
$19$ & $8 a a' H^2 \times \frac{\Delta \eta}{\Delta x^{4}} + 6 a^2 {a'}^2 H^4
\times \frac{\Delta \eta}{\Delta x^2} - 4 a^2 a' H^3 \times \frac1{\Delta x^2}$ \\
\hline
\end{tabular}
\caption{\footnotesize Each tabulated term must be multiplied by $-
\frac{\kappa^2}{64 \pi^4}$.}
\label{TL}
\end{table}
\noindent Because one of the propagators in the nonlocal diagrams (\ref{generic3pt}) 
might be undifferentiated, the functions $T^i(x;x')$ of equation (\ref{initialrep}) 
take the form,
\begin{equation}
T^i(x;x') \equiv T^i_N(x;x') + T^i_L(x;x') \!\times\! \ln(H^2 \Delta x^2) \; .
\label{TNTLdef}
\end{equation}
where $T^i_L(x;x')$ (given in Table~\ref{TL}) and $T^i_N(x;x')$ (given in 
Table~\ref{TN}) are functions of $a$, $a'$, $\Delta \eta \equiv \eta - \eta'$ and 
inverse powers of the Poincar\'e interval $\Delta x^2 \equiv \Vert \vec{x} - \vec{x}'
\Vert^2 - (\vert \eta - \eta'\vert - i \varepsilon)^2$. 
\begin{table}[H]
\setlength{\tabcolsep}{8pt}
\def\arraystretch{1.5}
\centering
\begin{tabular}{|@{\hskip 1mm }c@{\hskip 1mm }||c|}
\hline
$i$ & Coefficient $T^i_{N}(x;x')$ in expression (\ref{TNTLdef}) \\
\hline\hline
$1$ & $\!\!\frac{\frac{736}{5}}{\Delta x^8} \!-\! a a' \!H^2 [\frac{616 \Delta 
\eta^2}{\Delta x^8} \!+\! \frac{\frac{220}{3}}{\Delta x^6}]
\!-\! a^2 {a'}^2\! H^4 [\frac{96 \Delta \eta^4}{\Delta x^8} \!+\! 
\frac{\frac{812}{3} \Delta \eta^2}{\Delta x^6} \!+\! \frac{19}{\Delta x^4}]
\!-\! a^3 {a'}^3 \! H^6 [ \frac{64 \Delta \eta^4}{\Delta x^6} \!+\! 
\frac{22 \Delta \eta^2}{\Delta x^4}]\!\!$ \\
\hline
$2$ & $\!\!\frac{\frac{1952}{5}}{\Delta x^8} \!-\! a a' H^2 [\frac{416 \Delta 
\eta^2}{\Delta x^8} \!+\! \frac{\frac{128}{3}}{\Delta x^6}]
\!-\! a^2 {a'}^2 H^4 [\frac{\frac{112}{3} \Delta \eta^2}{\Delta x^6}
\!-\! \frac{56}{\Delta x^4}] \!+\! a^3 {a'}^3 H^6 [ \frac{32 \Delta \eta^4}{
\Delta x^6} \!+\! \frac{12 \Delta \eta^2}{\Delta x^4}]\!\!$ \\
\hline
$$ & $-184 [\frac{\frac{8}{5} \Delta \eta^2}{\Delta x^{10}} \!+\!
\frac{1}{\Delta x^8}] \!+\! 32 a a' H^2 [ \frac{36 \Delta \eta^4}{
\Delta x^{10}} \!+\! \frac{14 \Delta \eta^2}{\Delta x^8} \!-\! 
\frac{\frac{43}{3}}{\Delta x^6}]$  \\
$3$ & $- a^2 {a'}^2 H^4 [ \frac{288 \Delta \eta^4}{\Delta x^8} \!-\! 
\frac{\frac{1228}{3} \Delta \eta^2}{\Delta x^6} \!+\! \frac{145}{\Delta x^4}]
\!+\! 4 a^3 {a'}^3 H^6 [\frac{16 \Delta \eta^4}{\Delta x^6} \!+\! 
\frac{5 \Delta \eta^2}{\Delta x^4}]$ \\
$$ & $- 8 a H [\frac{232 \Delta \eta^3}{\Delta x^{10}} \!+\! 
\frac{203 \Delta \eta}{\Delta x^8}] \!+\! 4 a^2 a' H^3 
[ \frac{74 \Delta \eta^3}{\Delta x^8} \!-\! \frac{\frac{319}{3} \Delta \eta}{
\Delta x^6}] \!-\! a^3 {a'}^2 H^5 \!\times\! \frac{38 \Delta \eta}{\Delta x^4}$ \\
\hline
$5$ & $\frac{\frac{368}{5} \Delta \eta}{\Delta x^8} \!-\! 32 a a' H^2 
[ \frac{9 \Delta \eta^3}{\Delta x^8} \!-\! \frac{\frac43 \Delta \eta}{\Delta x^6}] 
\!+\! 4 a^2 {a'}^2 H^4 [ \frac{8 \Delta \eta^3}{\Delta x^6} \!-\!
\frac{\Delta \eta}{\Delta x^4}]$ \\
$$ & $+232 a H [\frac{2 \Delta \eta^2}{\Delta x^8} \!+\!
\frac{1}{\Delta x^6}] \!-\! 2 a^2 a' H^3
[\frac{\frac{28}{3} \Delta \eta^2}{\Delta x^6} \!-\! \frac{23}{\Delta x^4}] 
\!-\! a^3 {a'}^2 H^5 \!\times\! \frac{12 \Delta \eta^2}{\Delta x^4}$ \\
\hline
$7$ & $-\frac{\frac{92}{15}}{\Delta x^6} \!+\! a a' H^2
[\frac{24 \Delta \eta^2}{\Delta x^6} \!+\! \frac{\frac73}{\Delta x^4}] \!+\!
a^2 {a'}^2 H^4 [\frac{8 \Delta \eta^2}{\Delta x^4} \!-\! 
\frac{3}{\Delta x^2}] \!+\! a^3 {a'}^3 H^6 \!\times\! 
\frac{\Delta \eta^2}{\Delta x^2}$ \\
$$ & $- a H \!\times\! \frac{\frac{116}{3} \Delta \eta}{\Delta x^6} -
a^2 a' H^3 \!\times\! \frac{\frac{23}{3} \Delta \eta}{\Delta x^4} + 
a^3 {a'}^2 H^5 \!\times\! \frac{\Delta \eta}{\Delta x^2}$ \\
\hline
$9$ & $\!\!-16 [\frac{\frac{488}{5} \Delta \eta^2}{\Delta x^{10}} \!+\! 
\frac{61}{\Delta x^8}] \!+\! 16 a a' H^2 [\frac{88 \Delta \eta^4}{
\Delta x^{10}} \!-\! \frac{10 \Delta \eta^2}{\Delta x^8} \!-\! 
\frac{35}{\Delta x^6}]
\!+\! 4 a^2 {a'}^2 H^4 [\frac{8 \Delta \eta^4}{\Delta x^8} \!+\!
\frac{2 \Delta \eta^2}{\Delta x^6} \!-\! \frac{3}{\Delta x^4}]\!\!$ \\
\hline
$10$ & $\frac{\frac{976}{5} \Delta \eta}{\Delta x^8} \!-\! 16 a a' H^2 
[\frac{8 \Delta \eta^3}{\Delta x^8} \!-\! \frac{\frac{23}{3} \Delta 
\eta}{\Delta x^6}] \!-\! 2 a^2 {a'}^2 H^4 [ \frac{\frac{16}{3} \Delta 
\eta^3}{\Delta x^6} \!-\! \frac{23 \Delta \eta}{\Delta x^4}]$ \\
$$ & $+ a^3 {a'}^3 H^6 \!\times\! \frac{4 \Delta \eta^3}{\Delta x^4} \!+\! 
a H \!\times\! \frac{\frac{64}{3}}{\Delta x^6} \!+\! 8 a^2 a' H^3  
[ \frac{\frac{4}{3} \Delta \eta^2}{\Delta x^6} \!-\! \frac{5}{\Delta x^4}] \!-\!
a^3 {a'}^2 H^5 \!\times\! \frac{4 \Delta \eta^2}{\Delta x^4}$ \\  
\hline
$12$ & $-\frac{\frac{488}{15}}{\Delta x^6} \!+\! \frac{32}{3} a a' H^2 
[\frac{\Delta \eta^2}{\Delta x^6} \!-\! \frac1{\Delta x^4}] \!-\! a^2 {a'}^2 H^4 
[\frac{\frac{10}{3} \Delta \eta^2}{\Delta x^4} \!-\! \frac{8}{\Delta x^2}]
\!-\! a^3 {a'}^3 H^6 \!\times\! \frac{2 \Delta \eta^2}{\Delta x^2}$ \\
\hline
$13$ & $16 [\frac{336 \Delta \eta^4}{\Delta x^{12}} \!+\! 
\frac{336 \Delta \eta^2}{\Delta x^{10}} \!+\! \frac{63}{\Delta x^8}] + 4 a a' H^2
[\frac{336 \Delta \eta^4}{\Delta x^{10}} \!+\! \frac{868 \Delta \eta^2}{
\Delta x^8} \!+\! \frac{409}{\Delta x^6}]$ \\ 
$$ & $+ a^2 {a'}^2 H^4 [\frac{424 \Delta \eta^4}{\Delta x^8} \!+\!
\frac{144 \Delta \eta^2}{\Delta x^6} \!+\! \frac{557}{\Delta x^4}] \!-\! 
24 a^3 {a'}^3 H^6 [\frac{4 \Delta \eta^4}{\Delta x^6} \!-\! 
\frac{\Delta \eta^2}{\Delta x^4}]$ \\
\hline
$14$ & $-672 [\frac{\frac85 \Delta \eta^3}{\Delta x^{10}} \!+\! \frac{\Delta \eta
}{\Delta x^8}] \!-\! 8 a a' H^2 [\frac{30 \Delta \eta^3}{\Delta x^8} \!+\! 
\frac{\frac{107}{3} \Delta \eta}{\Delta x^6}] \!-\! 4 a^2 {a'}^2 H^4 
[\frac{16 \Delta \eta^3}{\Delta x^6} \!-\! \frac{7 \Delta \eta}{\Delta x^4}]$ \\
$$ & $-16 a H [\frac{4 \Delta \eta^2}{\Delta x^8} \!+\! \frac{\frac{35}{3}}{
\Delta x^6}] \!-\! 2 a^2 a' H^3 [\frac{\frac{68}{3} \Delta \eta^2}{\Delta x^6} \!+\!
\frac{137}{\Delta x^4}] \!+\! a^3 {a'}^2 H^5 \!\times\! \frac{16 \Delta \eta^2}{
\Delta x^4}$ \\
\hline
$16$ & $4 [\frac{\frac{84}{5} \Delta \eta^2}{\Delta x^8} \!+\! 
\frac{\frac{13}{3}}{\Delta x^6}] \!+\! a a' H^2 [\frac{24 \Delta \eta^2}{
\Delta x^6} \!+\! \frac{\frac{127}{3}}{\Delta x^4}] \!+\! a^2 {a'}^2 H^4 
[\frac{6 \Delta \eta^2}{\Delta x^4} \!+\! \frac{5}{\Delta x^2}]$ \\
$$ & $- a^3 {a'}^3 H^6 \!\times\! \frac{3 \Delta \eta^2}{\Delta x^2} \!-\! a H \!\times\!
\frac{28 \Delta \eta}{\Delta x^6} \!+\! a^2 a' H^3 \!\times\! \frac{\frac{49}{3} 
\Delta \eta}{\Delta x^4} \!+\! a^3 {a'}^2 H^5 \!\times\! \frac{\Delta \eta}{
\Delta x^2}$ \\
\hline
$18$ & $8 [ \frac{\frac{168}{5} \Delta \eta^2}{\Delta x^8} \!+\!
\frac{\frac{29}{3}}{\Delta x^6}] \!+\! 4 a a' H^2
[\frac{12 \Delta \eta^2}{\Delta x^6} \!-\! \frac{23}{\Delta x^4}] \!+\!
a^2 {a'}^2 H^4 \!\times\! \frac{10 \Delta \eta^2}{\Delta x^4}$ \\
\hline
$19$ & $-\frac{\frac{112}{5} \Delta \eta}{\Delta x^6} - 8 a a' H^2 \!\times\!
\frac{\Delta \eta}{\Delta x^4} - a^2 {a'}^2 H^4 \!\times\! \frac{\Delta \eta}{
\Delta x^2} + a H \!\times\! \frac{\frac{50}{3}}{\Delta x^4}$ \\
\hline
$21$ & $\frac{\frac{14}{5}}{\Delta x^4}$ \\
\hline
\end{tabular}
\caption{\footnotesize Each of the tabulated terms must be multiplied by 
$-\frac{\kappa^2}{64 \pi^4}$.}
\label{TN}
\end{table}

\subsubsection{Recovering the Renormalized Result}

We have developed a 4-step procedure for converting the unregulated expressions 
of Tables~\ref{TL} and \ref{TN} to fully renormalized results:
\begin{enumerate}
\item{Express $T^i_L(x;x')$ as derivatives acting on integrable functions;}
\item{Pass the multiplicative factor of $\ln(H^2 \Delta x^2)$ through the
derivatives;}
\item{Express the remainder $\Delta T^i_L(x;x')$ from step 2, plus 
$T^i_N(x;x')$, as derivatives acting on integrable functions and $1/\Delta x^4$;
and}
\item{Renormalize the factors of $1/\Delta x^4$ from step 3.}
\end{enumerate}
\noindent We will provide details and explain the rationale for each step 
below. As an example, we also implement the steps on the crucial coefficient
function $T^i(x;x')$ which controls corrections to the graviton mode function,
\begin{eqnarray}
\lefteqn{T^2_L(x;x') = -\frac{\kappa^2 \ln(H^2 \Delta x^2)}{64 \pi^4} \Biggl\{ 
a^2 {a'}^2 H^4 \Bigl[ -\frac{64 \Delta \eta^2}{\Delta x^6} - 
\frac{16}{\Delta x^4}\Bigr] } \nonumber \\
& & \hspace{6.5cm} + a^3 {a'}^3 H^6 \Bigl[-\frac{32 \Delta \eta^4}{\Delta x^6} - 
\frac{4}{\Delta x^2} \Bigr] \Biggr\} , \qquad \label{T2L1} \\
\lefteqn{T^2_N(x;x') = -\frac{\kappa^2}{64 \pi^4} \Biggl\{ \frac{\frac{1952}{3}}{
\Delta x^8} + a a' H \Bigl[ -\frac{416 \Delta \eta^2}{\Delta x^8} - 
\frac{\frac{128}{3}}{\Delta x^6}\Bigr] } \nonumber \\
& & \hspace{1.2cm} + a^2 {a'}^2 H^4 \Bigl[ -\frac{\frac{112}{3} \Delta \eta^2}{
\Delta x^6} + \frac{56}{\Delta x^4}\Bigr] + a^3 {a'}^3 H^6 \Bigl[\frac{32 \Delta 
\eta^4}{\Delta x^6} + \frac{12 \Delta \eta^2}{\Delta x^4} \Bigr] \Biggr\} .
\qquad \label{T2N1}
\end{eqnarray}

The rationale for {\bf Step 1} derives from the genesis of the logarithm 
contribution to the coefficient functions (\ref{TNTLdef}). The only way a 
factor of $\ln(H^2 \Delta x^2)$ can survive from one of the ghost or graviton 
propagators (\ref{ghostD4}-\ref{gravD4}) in a nonlocal contribution like 
(\ref{generic3pt}), is for all the derivatives to act on the other propagator. It 
is this differentiated propagator, times the scale factors from the vertices, 
that contribute to $T^i_L(x;x')$. Hence we must be able to express each 
$T^i_L(x;x')$ as a sum of terms involving scale factors times a number of 
derivatives acting on one of three integrable expressions,
\begin{equation}
\frac1{\Delta x^2} \qquad , \qquad \frac{\Delta \eta}{\Delta x^2} \qquad , 
\qquad \frac{\Delta \eta^2}{\Delta x^2} \; . \label{integrable3}
\end{equation}
Implementing {\bf Step 1} on expression (\ref{T2L1}) for $T^i_L(x;x')$
gives,
\begin{eqnarray}
\lefteqn{T^2_L(x;x') = -\frac{\kappa^2 \ln(H^2 \Delta x^2)}{64 \pi^4} \Biggl\{ 
a^2 {a'}^2 H^4 \!\times\! -8 \partial_0^2 \Bigl( \frac{1}{\Delta x^2} \Bigr) }
\nonumber \\
& & \hspace{3.5cm} + a^3 {a'}^3 H^6 \Bigl[-4 \partial_0^2 \Bigl( 
\frac{\Delta \eta^2}{\Delta x^2}\Bigr) \!+\! 20 \partial_0 \Bigl( 
\frac{\Delta \eta}{\Delta x^2}\Bigr) \!-\! \frac{16}{\Delta x^2} \Bigr] 
\Biggr\} . \qquad \label{T2L2}
\end{eqnarray}

In {\bf Step 2} we pass the overall multiplicative factor of $\ln(H^2 
\Delta x^2)$ through the derivatives to multiply the three integrable
functions (\ref{integrable3}). Implementing this on expression (\ref{T2L2})
for $T^i_{L}(x;x')$ gives,
\begin{eqnarray}
\lefteqn{T^2_L(x;x') = -\frac{\kappa^2 }{64 \pi^4} \Biggl\{ a^2 {a'}^2 H^4 
\!\times\! -8 \partial_0^2 \Bigl( \frac{\ln(H^2 \Delta x^2)}{\Delta x^2} \Bigr) 
+ a^3 {a'}^3 H^6 \Bigl[-4 \partial_0^2 } \nonumber \\
& & \hspace{0.5cm} \times \Bigl( \frac{\Delta \eta^2\ln(H^2 \Delta x^2)}{\Delta x^2}
\Bigr) + 20 \partial_0 \Bigl( \frac{\Delta \eta \ln(H^2 \Delta x^2)}{\Delta x^2}\Bigr) 
\!-\! \frac{16 \ln(H^2 \Delta x^2)}{\Delta x^2} \Bigr] \Biggr\} \qquad \nonumber \\
& & \hspace{1.5cm} -\frac{\kappa^2}{64 \pi^4} \Biggl\{ a^2 {a'}^2 H^4 \Bigl[- 
\frac{96 \Delta \eta^2}{\Delta x^6} - \frac{1}{\Delta x^4} \Bigr] + a^3 {a'}^3 H^6
\!\times\! -\frac{48 \Delta \eta^4}{\Delta x^6} \Biggr\} . \qquad \label{T2L3}
\end{eqnarray}
We can at this stage identify six integrable functions, with a factor of $2\pi i$
extracted for future convenience,
\begin{eqnarray} 
2 \pi i A_1 \equiv \frac{\ln(H^2 \Delta x^2)}{\Delta x^2} \qquad & , & \qquad
2 \pi i A_2 \equiv \frac{1}{\Delta x^2} \; , \qquad \label{A12def} \\
2 \pi i B_1 \equiv \frac{\Delta \eta\ln(H^2 \Delta x^2)}{\Delta x^2} \qquad & , & 
\qquad 2 \pi i B_2 \equiv \frac{\Delta \eta}{\Delta x^2} \; , \qquad 
\label{B12def} \\
2 \pi i C_1 \equiv \frac{\Delta \eta^2 \ln(H^2 \Delta x^2)}{\Delta x^2} 
\qquad & , & \qquad
2 \pi i C_2 \equiv \frac{\Delta \eta^2}{\Delta x^2} \; . \qquad \label{C12def}
\end{eqnarray}
Implementing {\bf Step 2} allows us to express $T^i_{L}(x;x')$ as a sum of
logarithm terms involving $A_1$, $B_1$ and $C_1$, which we term ``Group 1'', 
plus a ``remainder'' $\Delta T^i_{L}(x;x')$ which is devoid of logarithms. 
For example, expression (\ref{T2L3}) for $T^2_L(x;x')$ can be written as,
\begin{equation}
T^2_L = -\frac{i\kappa^2}{32 \pi^3} \Biggl\{ -8 a^2 {a'}^2 H^4 
\partial_0^2 A_1 \!-\! 4 a^3 {a'}^3 H^6 \Bigl[ \partial_0^2 C_1 \!-\! 5 
\partial_0 B_1 \!-\! 4 A_1\Bigr] \Biggr\} \!+\! \Delta T^2_{L} \; , 
\label{T2L4}
\end{equation}
where the remainder term is,
\begin{equation}
\Delta T^2_L(x;x') = -\frac{\kappa^2}{64 \pi^4} \Biggl\{ a^2 {a'}^2 H^4 \Bigl[- 
\frac{96 \Delta \eta^2}{\Delta x^6} - \frac{1}{\Delta x^4} \Bigr] + a^3 {a'}^3 
H^6 \!\times\! -\frac{48 \Delta \eta^4}{\Delta x^6} \Biggr\} . \label{T2L5}
\end{equation}

{\bf Step 3} begins by combining $T^i_{N}(x;x')$ with the remainder 
$\Delta T^i_{L}(x;x')$. For the case of $T^2(x;x')$ this consists of expressions 
(\ref{T2N1}) and (\ref{T2L5}),
\begin{eqnarray}
\lefteqn{T^2_N(x;x') + \Delta T^2_L(x;x') = -\frac{\kappa^2}{64 \pi^4} \Biggl\{ 
\frac{\frac{1952}{5}}{\Delta x^8} + a a' H \Bigl[ -\frac{416 \Delta \eta^2}{\Delta x^8} - 
\frac{\frac{128}{3}}{\Delta x^6}\Bigr] } \nonumber \\
& & \hspace{0.9cm} + a^2 {a'}^2 H^4 \Bigl[ -\frac{\frac{400}{3} \Delta \eta^2}{
\Delta x^6} + \frac{40}{\Delta x^4}\Bigr] + a^3 {a'}^3 H^6 \Bigl[-\frac{16 \Delta 
\eta^4}{\Delta x^6} + \frac{12 \Delta \eta^2}{\Delta x^4} \Bigr] \Biggr\} . \qquad 
\label{T2N2}
\end{eqnarray}
The resulting expressions can harbor ultraviolet divergences. One can see this by once
again extracting derivatives,
\begin{eqnarray}
\lefteqn{T^2_N(x;x') + \Delta T^2_L(x;x') } \nonumber \\
& & \hspace{-0.5cm} = -\frac{\kappa^2}{64 \pi^4} \Biggl\{ \partial^4 \Bigl( \frac{
\frac{61}{30}}{\Delta x^4} \Bigr) + a a' H \Bigl[ -\partial_0^2 \Bigl( \frac{
\frac{52}{3}}{\Delta x^4} \Bigr) + \partial^2 \Bigl( \frac{\frac{10}{3}}{\Delta x^4}
\Bigr) \Bigr] + a^2 {a'}^2 H^4  \nonumber \\
& & \hspace{-0.5cm} \times \Bigl[ -\partial_0^2 \Bigl(\frac{\frac{50}{3}}{\Delta x^2} 
\Bigr) \!+\! \frac{\frac{220}{3}}{\Delta x^4}\Bigr] \!+\! a^3 {a'}^3 H^6 \Bigl[-2 
\partial_0^2 \Bigl( \frac{\Delta \eta^2}{\Delta x^2}\Bigr) \!+\! 16 \partial_0 \Bigl(
\frac{\Delta \eta}{\Delta x^2} \Bigr) \!-\! \frac{12}{\Delta x^2} \Bigr] \Biggr\} . 
\qquad \label{T2N3}
\end{eqnarray}
In addition to the three integrable functions (\ref{integrable3}) one also encounters
logarithmically divergent factors of $1/\Delta x^4$. We consign the ultraviolet finite
factors of $A_2$, $B_2$ and $C_2$ to ``Group 2'' and reserve the factors of
$1/\Delta x^4$ for further analysis,
\begin{eqnarray}
\lefteqn{T^2_N \!+\! \Delta T^2_L = -\frac{i \kappa^2}{32 \pi^3} \Biggl\{-
\frac{50}{3} a^2 {a'}^2 H^4 \partial_0^2 A_2 \!-\! 2 a^3 {a'}^3 H^6 \Bigl[
\partial_0^2 C_2 \!-\! 8 \partial_0 B_2 \!+\! 6 A_2\Bigr] \Biggr\} } \nonumber \\
& & \hspace{-0.5cm} -\frac{\kappa^2}{64 \pi^4} \Biggl\{ \partial^4 \Bigl( \frac{
\frac{61}{30}}{\Delta x^4} \Bigr) \!-\! a a' H^2 \Bigl[ \frac{52}{3} \partial_0^2 
\!-\! \frac{10}{3} \partial^2\Bigr] \frac1{\Delta x^4} \!+\! \frac{220}{3} a^2 
{a'}^2 H^4 \Bigl( \frac1{\Delta x^4}\Bigr) \Biggr\} . \qquad \label{T2N4}
\end{eqnarray}

{\bf Step 4} is devoted to reducing and renormalizing the factor of 
$1/\Delta x^4$. In the dimensionally regulated computation it would derive 
from $1/\Delta x^{2D-4}$, from which we extract a factor of $\partial^2$,
\begin{equation}
\frac1{\Delta x^4} \longrightarrow \frac1{\Delta x^{2D-4}} = \frac{\partial^2}{
2 (D\!-\!3) (D\!-\!4)} \Bigl[\frac1{\Delta x^{2D-6}} \Bigr] \; . \label{newID1}
\end{equation}
The factor of $1/\Delta x^{2D-6}$ this produces is integrable in $D=4$ 
dimensions, so we could take the unregulated limit except for the explicit 
factor of $1/(D-4)$. We can add zero in the form of the propagator equation
for a massless scalar in flat space \cite{Onemli:2002hr,Onemli:2004mb},
\begin{eqnarray}
\lefteqn{\frac1{\Delta x^4} \longrightarrow \frac{\partial^2}{
2 (D\!-\!3) (D\!-\!4)} \Bigl[\frac1{\Delta x^{2D-6}} \Bigr] } \nonumber \\
& & \hspace{0cm} = \frac{\partial^2}{2 (D\!-\!3) (D\!-\!4)} \Bigl[\frac1{\Delta 
x^{2D-6}} - \frac{\mu^{D-4}}{\Delta x^{D-2}} \Bigr] + \frac{\mu^{D-4} 4 \pi^{\frac{D}2} 
i \delta^D(x \!-\! x')}{2 (D\!-\!3) (D\!-\!4) \Gamma(\frac{D}2 \!-\! 1)} \; . \qquad
\label{newID2}
\end{eqnarray}
We can take the unregulated limit of the nonlocal part of (\ref{newID2}),
\begin{equation}
\frac{\partial^2}{2 (D\!-\!3) (D\!-\!4)} \Bigl[\frac1{\Delta x^{2D-6}} - 
\frac{\mu^{D-4}}{\Delta x^{D-2}} \Bigr] \longrightarrow -\frac{\partial^2}{4} 
\Bigl[ \frac{\ln(\mu^2 \Delta x^2)}{\Delta x^2}\Bigr] \equiv -\frac{\partial^2}{4}
\Bigl[ 2\pi i A_3\Bigr] \; . \label{A3def}
\end{equation}
We call these ultraviolet finite terms ``Group 3A''.

The local divergence in the final part of expression (\ref{newID2}) requires 
explanation. First, recall from the generic expression (\ref{generic3pt}) that 
the nonlocal diagrams of Figure~\ref{diagrams} inherit a factor of $(a a')^{D-2}$
from the two vertices. Now, note from expressions (\ref{DeltaA}-\ref{Aprime}) 
that the first terms in the expansions of the dimensionally regulated ghost and 
graviton propagagors (\ref{ghostprop}-\ref{gravprop}) each have the form,
\begin{equation}
\frac{H^{D-2} \Gamma(\frac{D}2 \!-\! 1)}{(4 \pi)^{\frac{D}2}} \Bigl(
\frac{4}{y} \Bigr)^{\frac{D}{2}-1} = \frac{\Gamma(\frac{D}2 \!-\! 1)}{4 
\pi^{\frac{D}2}} \Bigl( \frac{1}{a a' \Delta x^2}\Bigr)^{\frac{D}2-1} \; .
\end{equation}
This means that the scale factors have no dependence on $D$, even 
in the fully regulated expression. On the other hand, the counterterms --- the 
final diagram in Figure~\ref{diagrams} --- each inherit a factor of $a^D$ 
from the universal $\sqrt{-g}$. Hence there is always a slight mismatch 
between the primitive divergences of the nonlocal diagrams and the 
counterterms which cancel them. This mismatch results in a local $\ln(a)$ in 
the renormalized, unregulated limit,
\begin{eqnarray}
\lefteqn{\frac{\mu^{D-4} 4 \pi^{\frac{D}2} i \delta^D(x \!-\! x')}{2 (D\!-\!3) 
(D\!-\!4) \Gamma(\frac{D}2 \!-\! 1)} - \frac{a^{D-4} \mu^{D-4} 4 \pi^{\frac{D}2}
i \delta^D(x \!-\! x')}{2 (D\!-\!3) (D\!-\!4) \Gamma(\frac{D}2 \!-\! 1)} }
\nonumber \\
& & \hspace{7cm} \longrightarrow - 2\pi^2 i \!\times\! \ln(a) 
\delta^4(x \!-\! x') \; . \qquad \label{renorm} 
\end{eqnarray}
These local terms are termed ``Group 3B''.

\subsubsection{The Schwinger-Keldysh Result}

The factors of $\delta^4(x - x')$ in Group 3B are already real 
and causal, however, the various nonlocal factors (\ref{A3def}) and 
(\ref{A12def}-\ref{C12def}) are not. This is because Feynman diagrams 
represent in-out matrix elements rather than true expectation values. The
Schwinger-Keldysh formalism \cite{Schwinger:1960qe,Mahanthappa:1962ex,
Bakshi:1962dv,Bakshi:1963bn,Keldysh:1964ud} is a diagrammtic technique for
computing true expectation values that is almost as simple to use as Feynman
diagrams. Expectation values of the graviton field obey effective field 
equations which are nonlocal but real and causal \cite{Chou:1984es,
Jordan:1986ug,Calzetta:1986ey}.

It is simple to convert an in-out 1PI $N$-point function such as the 
graviton self-energy to the Schwinger-Keldysh formalism. Without giving
the derivation, the rules are \cite{Ford:2004wc}:
\begin{itemize}
\item{Each spacetime point is assigned a polarity $\pm$.}
\item{One consequence is that every Feynman propagator $i\Delta(x;x')$ gives 
rise to four Schwinger-Keldysh propagators $i\Delta_{\pm \pm}(x;x')$. The
$++$ propagator is the same as the Feynman propagator and the $--$ propagator
is its complex conjugate. The $-+$ propagator is the free expectation of the
field at $x^{\mu}$ times the field at ${x'}^{\mu}$, and the $+-$ propagator 
is the free expectation value of the reverse-ordered product.}
\item{Another consequence is that each vertex has a $\pm$ polarity. The
$+$ vertices are the same as those of the in-out formalism while the $-$
vertices are complex cojugates.}
\item{A final consequence is that every in-out 1PI $N$-point function gives 
rise to $2^N$ $N$-point functions in the Schwinger-Keldysh formalism.} 
\item{The factor of $[\mbox{}^{\mu\nu} \Sigma^{\rho\sigma}](x;x')$ in the
linearized quantum Einstein equation (\ref{Einsteineqn}) is replaced by the
sum of $[\mbox{}^{\mu\nu} \Sigma^{\rho\sigma}_{++}](x;x')$, which is the same 
as the in-out result, and $[\mbox{}^{\mu\nu} \Sigma^{\rho\sigma}_{+-}](x;x')$.}
\item{On our simple background (\ref{geometry}), one can infer the result for
$[\mbox{}^{\mu\nu} \Sigma^{\rho\sigma}_{+-}](x;x')$ from that for 
$[\mbox{}^{\mu\nu} \Sigma^{\rho\sigma}](x;x')$ by dropping all the local
contributions of Group 3B, multiplying the nonlocal terms by $-1$,
and converting the coordinate interval $\Delta x^2$ from
\begin{equation}
\Delta x^2_{++}(x;x') \equiv \Bigl\Vert \vec{x} - \vec{x}' \Bigr\Vert^2 -
\Bigl( \vert \eta \!-\! \eta'\vert - i \varepsilon \Bigr)^2 \; , 
\label{Dx++}
\end{equation}
to
\begin{equation}
\Delta x^2_{+-}(x;x') \equiv \Bigl\Vert \vec{x} - \vec{x}' \Bigr\Vert^2 -
\Bigl( \eta \!-\! \eta' + i \varepsilon \Bigr)^2 \; . 
\label{Dx+-}
\end{equation}} 
\end{itemize}

Implementing these rules is straightforward. First, recall that the only 
dependence on the coordinate interval $\Delta x^2$ in the nonlocal results 
of Groups 1, 2, and 3A comes through the integrable functions $A_{1-3}$, 
$B_{12}$ and $C_{1-2}$, which were defined in expressions 
(\ref{A12def}-\ref{C12def}) and (\ref{A3def}). We can eliminate the factors 
of $1/\Delta x^2$ by extracting derivatives. For example, the $++$ and $+-$
versions of $2 \pi i \times A_1$ are,
\begin{equation}
2\pi i \!\times\! A_1 = \frac{\ln(H^2 \Delta x^2_{+\pm})}{\Delta x^2_{+\pm}}
= \frac{\partial^2}{8} \Bigl[ \ln^2(H^2 \Delta x^2_{+\pm}) \!-\! 2 \ln(H^2
\Delta x^2_{+\pm}) \Bigr] \; .
\end{equation}
Because the scale factors and the derivatives are identical in the $++$ and
$+-$ contributions, we need just consider differences of logarithms,
\begin{eqnarray}
\ln(H^2 \Delta x^2_{++}) - \ln(H^2 \Delta x^2_{+-}) & = & 2\pi i \!\times\!
\theta(\Delta \eta \!-\! r) \; , \\
\ln^2(H^2 \Delta x^2_{++}) - \ln^2(H^2 \Delta x^2_{+-}) & = & 4\pi i \!\times\!
\theta(\Delta \eta \!-\! r) \ln[H^2 (\Delta \eta^2 \!-\! r^2)] \; , \qquad 
\end{eqnarray} 
where $r \equiv \Vert \vec{x} - \vec{x}'\Vert$. The final reductions are,
\begin{eqnarray}
A_1 &\!\!\! \longrightarrow \!\!\!& +\frac{\partial^2}{4} 
\Biggl\{ \theta(\Delta \eta \!-\! r) \Bigl[ \ln[ H^2 (\Delta \eta^2 \!-\! r^2)]
\!-\! 1\Bigr] \Biggr\} \; , \qquad \label{ID4A} \\
B_1 &\!\!\! \longrightarrow \!\!\!& -\frac{\partial_0}{2} 
\Biggl\{ \theta(\Delta \eta \!-\! r) \ln[ H^2 (\Delta \eta^2 \!-\! r^2)]
\Biggr\} \; , \qquad \label{ID4B} \\
C_1 &\!\!\! \longrightarrow \!\!\!& +\frac{\partial_0^2}{4} 
\Biggl\{ \theta(\Delta \eta \!-\! r) (r^2 \!-\! \Delta \eta^2) \Bigl[
\ln[ H^2 (\Delta \eta^2 \!-\! r^2)] \!-\! 1\Bigr] \Biggr\} \nonumber \\
& & \hspace{5.5cm} + \frac12 \theta(\Delta \eta \!-\! r)
\ln[H^2 (\Delta \eta^2 \!-\! r^2)] \; , \qquad \label{ID4C} \\
A_2 &\!\!\! \longrightarrow \!\!\!& +\frac{\partial^2}{4} 
\Biggl\{ \theta(\Delta \eta \!-\! r) \Biggr\} \; , \qquad \label{ID4D} \\
B_2 &\!\!\! \longrightarrow \!\!\!& -\frac{\partial_0}{2} 
\Biggl\{ \theta(\Delta \eta \!-\! r) \Biggr\} \; , \qquad \label{ID4E} \\
C_2 &\!\!\! \longrightarrow \!\!\!& +\frac{\partial_0^2}{4} 
\Biggl\{ \theta(\Delta \eta \!-\! r) (r^2 \!-\! \Delta \eta^2) \Biggr\} 
+ \frac12 \theta(\Delta \eta \!-\! r) \; , \qquad \label{ID4F} \\ 
A_3 &\!\!\! \longrightarrow \!\!\!& +\frac{\partial^2}{4} 
\Biggl\{ \theta(\Delta \eta \!-\! r) \Bigl[ \ln[ \mu^2 (\Delta \eta^2 \!-\! r^2)]
\!-\! 1\Bigr] \Biggr\} \; . \qquad \label{ID4G} \\
\end{eqnarray}

\subsection{The 4-Point Contribution}

The previous discussion concerned the two nonlocal diagrams of 
Figure~\ref{diagrams}, and the local counterterms needed to renormalize them.
There are also finite local contributions from the 3rd diagram. It derives 
from the 42 4-graviton interactions given in equation (4.1) of 
\cite{Tsamis:1996qm}. One connects two of the graviton fields to the external
legs and then replaces the remaining two fields by graviton propagator. The 
procedure is tedious and we shall content ourselves with simply sketching it
and giving the final result. 

As an example we reduce the first of the 42 interactions, 
\begin{equation}
\mathcal{S}_1 \equiv \frac{\kappa^2}{32} \!\int\!\! d^Dx \, a^{D-2} h^2 
h_{,\theta} h^{,\theta} \; ,
\end{equation} 
where a comma denotes differentiation and the trace of the graviton field is
$h \equiv h^{\alpha}_{~\alpha} \equiv \eta^{\alpha\beta} h_{\alpha\beta}$. 
We first take the variational derivatives of the action integral with respect to 
$h_{\mu\nu}(x)$ and $h_{\rho\sigma}(x')$ as in expression (\ref{operatorexpr}),
\begin{eqnarray}
\lefteqn{\frac{i \delta^2 \mathcal{S}_1}{\delta h_{\mu\nu}(x) \delta 
h_{\rho\sigma}(x')} = \frac{\kappa^2}{32} \eta^{\mu\nu} \eta^{\rho\sigma} 
\Biggl\{ -\partial_{\theta} \Bigl[ 2 a^{D-2} h^{\alpha}_{~\alpha}(x) 
h^{\beta}_{~\beta}(x) \partial^{\theta} i \delta^D(x\!-\!x') \Bigr] } 
\nonumber \\
& & \hspace{0cm} + 4 a^{D-2} h^{\alpha}_{~\alpha}(x) h^{\beta}_{~\beta , 
\theta}(x) \partial^{\theta} i \delta^D(x \!-\! x') \!-\! 4 \partial^{\theta} 
\Bigl[ a^{D-2} h^{\alpha}_{~\alpha}(x) h^{\beta}_{~\beta , \theta}(x) 
i \delta^D(x \!-\! x') \Bigr] \nonumber \\
& & \hspace{5.5cm} + 2 a^{D-2} h^{\alpha}_{~\alpha , \theta}(x)
h^{\beta}_{~\beta ,\theta}(x) \Big{]} i \delta^D(x\!-\!x') \Biggr\} . \qquad 
\end{eqnarray}
Now take the expectation value of the $T^*$-ordered product, which amounts to 
replacing the remaining two graviton fields of each term by the appropriate 
coincident (and sometimes differentiated) propagator,
\begin{eqnarray}
\lefteqn{\Bigl\langle \Omega \Bigl\vert T^*\Bigl[\frac{i \delta^2 \mathcal{S}_1}{
\delta h_{\mu\nu}(x) \delta h_{\rho\sigma}(x')} \Bigr] \Bigr\vert \Omega 
\Bigr\rangle = \frac{\kappa^2}{32} \eta^{\mu\nu} \eta^{\rho\sigma} \Biggl\{ 
-\partial_{\theta} \Biggl[ 2 a^{D-2} \!\times\! i\!\Bigl[\mbox{}^{\alpha}_{~\alpha} 
\Delta^{\beta}_{~\beta}\Bigr]\!(x;x) } \nonumber \\
& & \hspace{-0.7cm} \times \partial^{\theta} i \delta^D(x\!-\!x') \Biggr] \!+\! 
4 a^{D-2} \!\times\! \partial'_{\theta} i\!\Bigl[\mbox{}^{\alpha}_{~\alpha} 
\Delta^{\beta}_{~\beta}\Bigr]\!(x;x')\Bigl\vert_{x'=x} \!\!\!\!\!\!\!\!\times 
\partial^{\theta} i \delta^D\!(x \!-\! x') \!-\! 4 \partial^{\theta} \Biggl[ 
a^{D-2}  \nonumber \\
& & \hspace{-0.7cm} \times i\delta^D\!(x \!-\! x') \!\times\! \partial'_{\theta} 
i\!\Bigl[\mbox{}^{\alpha}_{~\alpha} \Delta^{\beta}_{~\beta} \Bigr]\!(x;x') \!\Biggr] 
\!\!+\! 2 a^{D-2} i \delta^D\!(x\!-\!x') \!\times\! \partial_{\theta} 
\partial'^{\theta} i\!\Bigl[\mbox{}^{\alpha}_{~\alpha} \Delta^{\beta}_{~\beta}
\Bigr]\!(x;x') \!\Biggr\} . \qquad
\end{eqnarray}
Finally, we express the tensor structure using the 21 basis tensors of 
Group 1,
\begin{eqnarray}
\eta^{\mu\nu} \eta^{\rho\sigma} & = & \Bigl( \overline{\eta}^{\mu\nu} \!-\!
\delta^{\mu}_{~0} \delta^{\nu}_{~0} \Bigr) \Bigl(\overline{\eta}^{\rho\sigma}
\!-\! \delta^{\rho}_{~0} \delta^{\sigma}_{~0}\Bigr) \; , \\
& = & \Bigl[\mbox{}^{\mu\nu} \mathcal{D}_1^{\rho\sigma}\Bigr] \!-\!
\Bigl[\mbox{}^{\mu\nu} \mathcal{D}_3^{\rho\sigma}\Bigr] \!-\!
\Bigl[\mbox{}^{\mu\nu} \mathcal{D}_4^{\rho\sigma}\Bigr] \!+\!
\Bigl[\mbox{}^{\mu\nu} \mathcal{D}_{13}^{\rho\sigma}\Bigr] \; . \qquad 
\end{eqnarray}

The coincidence limits of the three propagators which appear in the graviton
propagator (\ref{gravprop}) are,
\begin{eqnarray} 
i\Delta_A(x;x) = k \Bigl[-\pi {\rm cot}\Bigl( \frac{\pi D}{2}\Bigr) \!+\! 2
\ln(a) \Bigr] \quad & , & \quad i\Delta_B(x;x) = -\frac{k}{D\!-\!2} \; , 
\qquad \\
i\Delta_C(x;x) = \frac{k}{(D\!-\!2)(D\!-\!3)} \quad & , & \quad k \equiv 
\frac{H^{D-2}}{(4\pi)^{\frac{D}2}} \frac{\Gamma(D\!-\!1)}{\Gamma(\frac{D}2)}
\; . \qquad 
\end{eqnarray}
Note that only the undifferentiated $A$-type propagator is ultraviolet
divergent in dimensional regularization. The undifferentiated $A$-type propagator
is also the only way to get a factor of $\ln(a)$. First derivatives of coincident 
propagators are all finite,
\begin{equation}
\partial_{\alpha} i\Delta_A(x;x') \Bigl\vert_{x'=x} = a H k \delta^0_{~\alpha}
\;\; , \;\; \partial_{\alpha} i\Delta_B(x;x') \Bigl\vert_{x'=x} = 0 =
\partial_{\alpha} i\Delta_C(x;x') \Bigl\vert_{x'=x} \; .
\end{equation}
Mixed second derivatives are also finite,
\begin{eqnarray}
\partial_{\alpha} \partial'_{\beta} i\Delta_A(x;x') \Bigl\vert_{x'=x} & = &
-\Bigl( \frac{D\!-\!1}{D} \Bigr) k H^2 g_{\alpha\beta} \; , \\
\partial_{\alpha} \partial'_{\beta} i\Delta_B(x;x') \Bigl\vert_{x'=x} & = &
\frac{1}{D} k H^2 g_{\alpha\beta} \; , \\
\partial_{\alpha} \partial'_{\beta} i\Delta_C(x;x') \Bigl\vert_{x'=x} & = &
-\frac{2}{D (D \!-\! 2)} k H^2 g_{\alpha\beta} \; .
\end{eqnarray}

Note that all primitive contributions have factors of $a^{D-2}$, $a^{D-1}$
or $a^D$. The counterterms which absorb ultraviolet divergences possess 
the very same dependence on $a$ so renormalization engenders no finite factors 
of $\ln(a)$ the way it did for the nonlocal diagrams of expression 
(\ref{renorm}). It does produce factors of $\ln(H/\mu)$ but we report only the
$\ln(a)$ contributions to $i T^2(x;x')$,
\begin{equation}
\frac{\kappa^2 \ln(a)}{32 \pi^2} \!\times\! 8 a^2 H^2 (\partial_0 \!+\! 2 a H) 
\partial_0 \delta^4(x \!-\! x') \; .
\end{equation}
We call this class of contributions ``Group 4''.

\section{The Effect on Gravitational Radiation}

The purpose of this section is to solve (\ref{Einsteineqn}) for $T^{\mu\nu}_{~\rm 
lin}(x) = 0$ to derive one loop corrections to the graviton mode function. We begin
by deriving an equation for the graviton mode function, then consider the one loop 
response to various sources. To simplify the computation we specialize the various
source terms to late times and small wave numbers. Finally, we compute the source
integrals and read off the late time result for the one loop correction.

Dynamical gravitons correspond to zero stress tensor and the wave form,
\begin{equation}
h_{\mu\nu}(x;\vec{k}) = \epsilon_{\mu\nu}(\vec{k}) u(\eta,k) e^{i \vec{k} \cdot 
\vec{x}} \qquad , \qquad 0= \epsilon_{\mu 0} = \epsilon_{ij} k_j = \epsilon_{ii} 
\; . \label{gravform}
\end{equation}
The graviton polarization tensors $\epsilon_{\mu\nu}(\vec{k})$ is identical to the
usual ones from flat space. From their properties, and the $3+1$ decomposition 
(\ref{D00}-\ref{Dij}) of the gauge-fixed kinetic operator, we find,
\begin{equation}
\mathcal{D}^{\mu\nu\rho\sigma} h_{\rho\sigma}(x;\vec{k}) = 
\epsilon^{\mu\nu}(\vec{k}) e^{i \vec{k} \cdot \vec{x}} \!\times\! -\frac12 a^2 
\Bigl[ \partial_0^2 \!+\! 2 H a \partial_0 \!+\! k^2\Bigr] u(\eta,k) \; . 
\label{gravlhs}
\end{equation}
The graviton wave form (\ref{gravform}) implies $0 = h_{00} = h_{0i} = h_{ij,j}$.
Using these relations in the $3+1$ decomposition (\ref{Sigma00}-\ref{Sigmaij}) of 
the quantum source term implies,
\begin{eqnarray}
\lefteqn{\int \!\! d^4x' \Bigl[\mbox{}^{\mu\nu} \Sigma^{\rho\sigma}\Bigr](x;x') 
h_{\rho\sigma}(x';\vec{k}) } \nonumber \\
& & \hspace{3cm} = \epsilon^{\mu\nu}(\vec{k}) e^{i \vec{k} \cdot \vec{x}}
\!\times\!\! \int \!\! d^4x' \, iT^2_{SK}(x;x') u(\eta',k) e^{-i \vec{k} \cdot
(\vec{x} - \vec{x}')} \; . \qquad \label{gravrhs}
\end{eqnarray}
Equating the left hand side (\ref{gravlhs}) with the right hand side (\ref{gravrhs})
gives the graviton mode equation,
\begin{equation}
-\frac12 a^2 \Bigl[ \partial_0^2 \!+\! 2 H a \partial_0 \!+\! k^2\Bigr] u(\eta,k) =
\int \!\! d^4x' \, iT^2_{SK}(x;x') u(\eta',k) e^{-i \vec{k} \cdot
\Delta \vec{x}} \; . \label{gravmodeeqn}
\end{equation} 

The graviton mode equation (\ref{gravmodeeqn}) is valid to all orders, but we only
possess one loop results for the coefficient function $T^2_{SK}(x;x')$. Hence we
must solve the equation perturbatively. The 0th order mode function, and its
asymptotic late time expansion, are, 
\begin{equation}
u_0(\eta,k) = \frac{H}{\sqrt{2 k^3}} \Bigl[1 - \frac{i k}{H a} \Bigr] 
\exp\Bigl[\frac{i k}{H a}\Bigr] \longrightarrow \frac{H}{\sqrt{2 k^3}} 
\Bigl[1 + \frac{k^2}{2 H^2 a^2} + \dots\Bigr] . \label{uzero}
\end{equation}
The one loop correction we seek obeys the equation,
\begin{equation}
-\frac12 a^2 \Bigl[ \partial_0^2 \!+\! 2 H a \partial_0 \!+\! k^2\Bigr] u_1(\eta,k)
= \int \!\! d^4x' \, iT^2_{SK}(x;x') u_0(\eta',k) e^{-i \vec{k} \cdot \Delta \vec{x}} 
\equiv S(\eta) \; . \label{grav1loop}
\end{equation} 
Finally, let us consider the relation between possible late time behaviors for the
quantum source $S(\eta)$ and the corresponding asymptotic late time form of the one 
loop mode function,
\begin{eqnarray}
S = a^4 \ln(a) \Longrightarrow u_1 \longrightarrow -\frac{\ln^2(a)}{3 H^2} & , &
S = a^4 \Longrightarrow u_1 \longrightarrow -\frac{2 \ln(a)}{3 H^2} \; , \qquad 
\label{a4} \\
S = a^3 \ln(a) \Longrightarrow u_1 \longrightarrow +\frac{\ln(a)}{H^2 a} & , &
S = a^3 \Longrightarrow u_1 \longrightarrow +\frac{1}{H^2 a} \; , \qquad 
\label{a3} \\
S = a^2 \ln(a) \Longrightarrow u_1 \longrightarrow +\frac{\ln(a)}{H^2 a^2} & , &
S = a^2 \Longrightarrow u_1 \longrightarrow +\frac{1}{H^2 a^2} \; . \label{a2}
\end{eqnarray}

Let us start with the local contributions of Groups 3B and 4,
\begin{eqnarray}
S_{3B} &\!\!\! = \!\!\!& \frac{\kappa^2 \ln(a)}{32 \pi^2} \Biggl\{ -\frac{61}{30} 
\partial^4 u_0 + a H^2 \Bigl[\frac{52}{3} \partial_0^2 \!-\! \frac{10}{3}
\partial^2\Bigr] (a u_0) - \frac{220}{3} a^4 H^4 u_0\Biggr\} , \qquad
\label{Slocal} \\
S_{4} &\!\!\! = \!\!\!& \frac{\kappa^2 \ln(a)}{32 \pi^2} \Biggl\{8 a^2 H^2
\partial_0^2 u_0 \!+\! 16 a^3 H^3 \partial_0 u_0 \Biggr\} , \label{Slocal4pt}
\end{eqnarray}
where $\partial^2 \rightarrow -(\partial_0^2 + k^2)$. Expression (\ref{Slocal4pt})
was studied previously when making the Hartree approximation \cite{Mora:2013ypa},
but it is simple enough to evaluate both local contributions
(\ref{Slocal}-\ref{Slocal4pt}) using relations such as,
\begin{eqnarray}
\partial_0 u_0(\eta,k) = \frac{H}{\sqrt{2 k^3}} \!\times\! k^2 \eta e^{-ik \eta} 
& , & \partial_0^2 u_0(\eta,k) = \frac{H}{\sqrt{2 k^3}} \Bigl(k^2 \!-\! i k^3 
\eta\Bigr) e^{-ik \eta} \; , \qquad \label{d0u0A} \\
(\partial_0^2 \!+\! k^2)^2 u_0(\eta,k) = 0 & , & (\partial_0^2 \!+\! k^2) 
\Bigl( a u_0(\eta,k)\Bigr) = 2 a^3 H^2 u_0(\eta,k) \; . \qquad \label{d0u0B}
\end{eqnarray}
Hence (\ref{Slocal}-\ref{Slocal4pt}) are,
\begin{eqnarray}
S_{3B}(\eta) & = & \frac{\kappa^2 H^4}{32 \pi^2} \!\times\! a^4 \ln(a) 
\!\times\! \Bigl[-32 \!-\! \frac{52}{3} \frac{k^2}{a^2 H^2} \Bigr] u_0(\eta,k)
\; , \qquad \label{gravlocal} \\
S_{4}(\eta) & = & \frac{\kappa^2 H^4}{32 \pi^2} \!\times\! a^4 \ln(a) 
\!\times\! \Bigl[0 \!-\! \frac{8 k^2}{a^2 H^2} \Bigr] u_0(\eta,k) \; . \qquad 
\label{gravlocal4pt} 
\end{eqnarray}
However, to recover the leading late time behavior we can set $\partial_0^2 + k^2$ 
to just $\partial_0^2$, and $u_0(\eta,k)$ to just $u_0(0,k)$,
\begin{eqnarray}
S_{3B}(\eta) & \longrightarrow & \frac{\kappa^2 H^4}{4 \pi^2} 
\!\times\! u_0(0,k) \!\times\! -4 a^4 \ln(a) \; , \qquad \label{gravlocallate} \\
S_{4}(\eta) & \longrightarrow & \frac{\kappa^2 H^4}{4 \pi^2} 
\!\times\! u_0(0,k) \!\times\! 0 \; . \qquad \label{gravlocal4ptlate} 
\end{eqnarray}

Old results \cite{Prokopec:2002uw,Duffy:2005ue} could be exploited to evaluate 
the nonlocal contributions from Groups 1, 2 and 3A 
exactly, the same way we did with the local contribution (\ref{gravlocal}).
However, if we are only interested in the leading behavior at late times, as
in expression (\ref{gravlocallate}), we can replace $u_0(\eta',k) e^{-i \vec{k}
\cdot \Delta \vec{x}}$ with $u_0(0,k)$, which means that $\partial^2$
degenerates to just $-\partial_0^2$. We also change $\partial_0$ to $-\partial_0'$
and partially integrate to act on any primed scale factors, or give zero if 
there are no primed scale factors. The source integrals for Groups 1, 2 and 
3A are,
\begin{eqnarray}
\lefteqn{ S_{1} \longrightarrow \frac{\kappa^2}{32 \pi^3} \!\times\!
u_0(0,k) \!\times\!\! \int \!\! d^4x' \, \theta(\Delta \eta \!-\! r) \Biggl\{
-144 a^3 {a'}^5 H^8 \ln[H^2 (\Delta \eta^2 \!-\! r^2)] } \nonumber \\
& & \hspace{1cm} + \Bigl[240 a^2 {a'}^6 H^8 \!+\! 48 a^3 {a'}^5 H^8 \!-\! 360
a^3 {a'}^7 H^{10} (r^2 \!-\! \Delta \eta^2) \Bigr] \nonumber \\
& & \hspace{7.3cm} \times \Bigl[ \ln[H^2 (\Delta \eta^2 \!-\! r^2)] \!-\! 1\Bigr] 
\Biggr\} , \label{S5I} \qquad \\
\lefteqn{ S_{2} \longrightarrow \frac{\kappa^2}{32 \pi^3} \!\times\!
u_0(0,k) \!\times\!\! \int \!\! d^4x' \, \theta(\Delta \eta \!-\! r) \Biggl\{
500 a^2 {a'}^6 H^8 \!-\! 72 a^3 {a'}^5 H^8 } \nonumber \\
& & \hspace{7.3cm} - 180 a^3 {a'}^7 H^{10} (r^2 \!-\! \Delta \eta^2) \Biggr\} ,
\label{S6I} \qquad \\
\lefteqn{ S_{3A} \longrightarrow \frac{\kappa^2}{32 \pi^3} \!\times\!
u_0(0,k) \!\times\!\! \int \!\! d^4x' \, \theta(\Delta \eta \!-\! r) \Biggl\{
\Bigl[930 a {a'}^7 H^8 \!-\! 550 a^2 {a'}^6 H^8\Bigr] } \nonumber \\
& & \hspace{7.5cm} \times \Bigl[ \ln[\mu^2 (\Delta \eta^2 \!-\! r^2)] \!-\! 1
\Bigr] \Biggr\} . \qquad \label{S7I}
\end{eqnarray}
  
The next step is to perform the spatial integrations,
\begin{eqnarray}
\lefteqn{\int \!\! d^3x' \, \theta(\Delta \eta \!-\! r) 
= \frac{4\pi}3 \Delta \eta^3 \theta(\Delta \eta) \; , } \\
\lefteqn{\int \!\! d^3x' \, \theta(\Delta \eta \!-\! r) \ln[H^2 (\Delta \eta^2 
\!-\! r^2)] 
= \frac{8\pi}3 \Bigl[\ln(2 H \Delta \eta) \!-\! \frac43\Bigr] 
\Delta \eta^3 \theta(\Delta \eta) \; , } \\
\lefteqn{\int \!\! d^3x' \, \theta(\Delta \eta \!-\! r) (r^2 \!-\! \Delta \eta^2)
= -\frac{8\pi}{15} \Delta \eta^5 \theta(\Delta \eta) \; , } \\
\lefteqn{\int \!\! d^3x' \, \theta(\Delta \eta \!-\! r) (r^2 \!-\! \Delta \eta^2) 
\ln[H^2 (\Delta \eta^2 \!-\! r^2)] } \nonumber \\
& & \hspace{5.4cm} = -\frac{8\pi}{15} \Bigl[2 \ln(2 H \Delta \eta) \!-\! 
\frac{31}{15}\Bigr] \Delta \eta^5 \theta(\Delta \eta) \; . \qquad
\end{eqnarray}
We also change variables from $\eta'$ to $a' =-1/H\eta'$,
\begin{equation}
\int_{\eta_i}^{\eta} \!\!\!\! d\eta' = \int_{1}^{a} \!\! \frac{da'}{H {a'}^2} \; , 
\end{equation}
where the initial time $\eta_i = -1/H$ corresponds to unit scale factor. Note 
also that $a' H \Delta \eta = 1 - a'/a$. These reductions allow us to express
(\ref{S5I}-\ref{S7I}) as,
\begin{eqnarray}
\lefteqn{S_{1} \longrightarrow \frac{\kappa^2 H^4}{4 \pi^2} 
u_0(0,k) \!\!\! \int_1^a \!\!\!\! da'\Biggl\{ \!-48 a^3 \Bigl(\!1 \!-\! \frac{a'}{a}
\Bigr)^{\!3} \!\Bigl[ \ln(2 H \Delta \eta) \!-\! \frac43\Bigr] \!+\! \Bigl[ 80 a^2 a' 
\!+\! 16 a^3 \Bigr] } \nonumber \\
& & \hspace{0.5cm} \times \Bigl(1 \!-\! \frac{a'}{a}\Bigr)^3 \Bigl[\ln(2 H \Delta \eta) 
\!-\! \frac{11}{6}\Bigr] + 48 a^3 \Bigl(1 \!-\! \frac{a'}{a}\Bigr)^5 \Bigl[ \ln(2 H 
\Delta \eta) \!-\! \frac{23}{15} \Bigr] \Biggr\} , \qquad \label{S5II} \\
\lefteqn{S_{2} \longrightarrow \frac{\kappa^2 H^4}{4 \pi^2}
u_0(0,k) \!\!\! \int_1^a \!\!\!\! da'\Biggl\{ \Bigl[ \frac{250}{3} a^2 a' \!-\! 12 a^3
\Bigr] \Bigl(\!1 \!-\! \frac{a'}{a}\Bigr)^{3} \!\!+\! 12 a^3 \Bigl(1 \!-\! 
\frac{a'}{a}\Bigr)^5 \Biggr\} , \qquad} \label{S6II} \\
\lefteqn{S_{3A} \longrightarrow \frac{\kappa^2 H^4}{4 \pi^2}
u_0(0,k) \!\!\! \int_1^a \!\!\!\! da' \Bigl[310 a {a'}^2 \!-\! \frac{550}{3} a^2 a'
\Bigr] \Bigl(\!1 \!-\! \frac{a'}{a}\Bigr)^{3} \!\Bigl[ \ln(2 \mu \Delta \eta) \!-\! 
\frac{11}{6}\Bigr] . \qquad} \label{S7II}
\end{eqnarray}

It remains to perform the temporal integrations. These can be done exactly but
we give only the coefficients of the leading $a^4 \ln(a)$ dependence, 
\begin{eqnarray}
S_{1} & \longrightarrow & \frac{\kappa^2 H^4}{4 \pi^2} \!\times\!
u_0(0,k) \!\times\! -4 a^4 \ln(a) \; , \qquad \label{S5III} \\
S_{2} & \longrightarrow & \frac{\kappa^2 H^4}{4 \pi^2} \!\times\!
u_0(0,k) \!\times\! 0 \; , \qquad \label{S6III} \\
S_{3A} & \longrightarrow & \frac{\kappa^2 H^4}{4 \pi^2} \!\times\!
u_0(0,k) \!\times\! 4 a^4 \ln(a) \; . \qquad \label{S7III}
\end{eqnarray}
Combining these with the local results (\ref{gravlocallate}-\ref{gravlocal4ptlate})
we see that the late time asymptotic form of the total source is,\footnote{It is 
interesting to note that the terms of Groups 3A and 3B, which 
ultimately descended from factors of $1/\Delta x^4$, cancel one another at 
leading order. It is also interesting that the nonzero contribution from 
Group 1 derives entirely from $A_1$ because the $B_1$ and $C_1$ 
contributions also cancel.}
\begin{equation}
S(\eta) \longrightarrow \frac{\kappa^2 H^4}{4 \pi^2} \!\times\!
u_0(0,k) \!\times\! -4 a^4 \ln(a) \; . \label{sourcelate}
\end{equation}
In view of expression (\ref{a4}) it follows that the asymptotic form of the one 
loop correction to the mode function is,
\begin{equation}
u_1(\eta,k) \longrightarrow \frac{\kappa^2 H^2}{4\pi^2} \!\times\! u_0(0,k)
\!\times\! \frac43 \ln^2(a) \; . \label{lateu1}
\end{equation}

\section{Epilogue}

In section 2 of this paper we give a 4-step procedure for converting an old, 
unregulated computation \cite{Tsamis:1996qk} of the single graviton loop 
contribution to the self-energy $-i[\mbox{}^{\mu\nu} \Sigma^{\rho\sigma}](x;x')$ 
to the fully renormalized result, with only one additional computation needed for 
the 4-point contribution. The technique laid out in Section 2 has been used to 
obtain results for all 21 of the coefficient functions $T^i(x;x')$ in the 
representation (\ref{initialrep}), and these will be reported in a subsequent 
work \cite{Tan:2022xpn}. Here we give only the coefficient $T^2(x;x')$ which enters 
the equation (\ref{gravmodeeqn}) for the mode function $u(\eta,k)$ of plane 
wave gravitons (\ref{gravform}). In section 3 we applied this result to derive 
the asymptotic late time form (\ref{lateu1}) for one loop corrections to 
$u(\eta,k)$.

There are two striking features about our result (\ref{lateu1}).
First, it grows with time. We saw in section 3 that this growth derives entirely
from the ``tail'' part of the graviton propagator, which is associated with the
continual production of gravitons with Hubble-scale momenta. The physical 
interpretation is that the spin-spin coupling allows our test graviton to 
remain in interaction with these gravitons, long after its own kinetic energy 
has red-shifted to insignificance. The interaction produces a growth in the
amplitude for the simple reason that scattering between very long wavelength 
and shorter wavelength particles tends to transfer momentum from the latter to 
the former. Even though the loop counting parameter $\kappa^2 H^2 \ltwid 
10^{-10}$ is minuscule for actual inflation, this growth must eventually cause 
perturbation theory to break down if inflation persists long enough.

The second striking feature of our result is that it implies a potentially 
observable one loop correction to the tensor power spectrum, 
\begin{equation}
\Delta_{h}^2(k) \simeq \frac{16 \hbar G H^2(t_k)}{\pi c^3} \Biggl\{ 1 + 
\frac{16 \hbar G H^2(t_k)}{3 \pi c^3} \!\times\! N_k^2 + O(G^2) \Biggr\} \; ,
\label{power}
\end{equation}
where $H(t_k)$ is the value of the Hubble parameter at horizon crossing 
(defined by $a(t_k) H(t_k) = k$) and $N_k$ is the number of e-foldings 
from horizon crossing to the end of inflation.\footnote{Note that our 
result obeys the bound on loop corrections to the primordial power spectrum 
derived by the late Steven Weinberg \cite{Weinberg:2005vy,Weinberg:2006ac}.} 
No one knows how small the tree order power spectrum is, but if it is near 
the current upper bound of $10^{-10}$ \cite{Planck:2018vyg} then the large 
factor of $N_k^2$ in (\ref{power}) might allow the one loop correction to 
be resolved in the far future, after the full development of 21cm cosmology 
\cite{Loeb:2003ya,Furlanetto:2006jb,Masui:2010cz}.

The great uncertainty in all this is the gauge issue. The graviton
propagator requires gauge fixing, and how one accomplishes that can affect
loop corrections. This is well known in flat space. For example, the 
numerical coefficient $\frac{61}{30}$ in the first term of equation 
(\ref{Slocal}) derives from flat space. When the graviton self-energy is 
re-computed in the most general Poincar\'e invariant gauge ($\mathcal{L} = 
-\frac1{2 \alpha} F_{\mu} F^{\mu}$, with $F_{\mu} = \partial^{\nu} 
h_{\mu\nu} - \frac{\beta}{2} \partial_{\mu} h$), this numerical coefficient 
becomes \cite{Capper:1979ej},
\begin{equation}
\frac{61}{30} \longrightarrow \frac{1}{2} \alpha^2 + \frac{3}{4} \alpha 
- \frac{43}{60} + \frac{1}{6} \frac{(\alpha \!-\! 3)^2}{(\beta \!-\! 2)^4} 
- \frac{7}{4} \frac{(\alpha \!-\! 3)}{(\beta \!-\! 2)^3} - \frac{1}{12} 
\frac{(\alpha \!-\! 51)}{(\beta \!-\! 2)^2} + \frac{1}{12} \frac{(9 \alpha 
\!-\! 11)}{(\beta \!-\! 2)} \; . \label{Coef}
\end{equation}
This particular term makes zero correction to $u(\eta,k)$, and the gauge 
dependence of the distinctly de Sitter terms which produce (\ref{lateu1}) 
is not known. Computations on de Sitter background are so difficult that 
all but one \cite{Glavan:2015ura} of the graviton loop results 
\cite{Tsamis:1996qm,Tsamis:1996qk,Tsamis:2005je,Miao:2005am,Kahya:2007bc,
Miao:2012bj,Leonard:2013xsa,Boran:2014xpa,Boran:2017fsx,Glavan:2020gal}
so far obtained have been made using the simplest gauge \cite{Tsamis:1992xa,
Woodard:2004ut}. The single exception \cite{Glavan:2015ura} was a 
re-computation of the vacuum polarization \cite{Leonard:2013xsa} in a 
1-parameter family of de Sitter invariant gauges \cite{Mora:2012zi}.
When that result was used to compute quantum gravitational corrections
the photon field strength \cite{Glavan:2016bvp}, the same asymptotic time
dependence was found as for the simple, de Sitter breaking gauge 
\cite{Wang:2014tza}, but with a different numerical coefficient, and no
dependence at all on the gauge parameter. One must therefore assume a
slight gauge dependence in cosmological solutions to the naive effective
field equations.

John Donoghue has demonstrated how general relativity can be used as a 
low energy effective field theory on flat space background to derive
quantum gravitational corrections to the long-range potentials carried 
by massless particles such as photons and gravitons \cite{Donoghue:1993eb,
Donoghue:1994dn}. The method is first to compute the scattering amplitude 
between two massive particles that exchange the massless field, then 
infer the exchange potential using inverse scattering theory. That is 
how gauge independent, graviton loop corrections were derived for the 
Newtonian \cite{Bjerrum-Bohr:2002fji,Bjerrum-Bohr:2002gqz} and Coulomb  
\cite{Bjerrum-Bohr:2002aqa} potentials. 

It has recently been shown that Donoghue's S-matrix technique can be
applied directly to produce gauge independent effective field equations, 
without computing cosmologically unobservable scattering amplitudes 
\cite{Miao:2017feh}. The new procedure employs position space 
versions of crucial identities Donoghue and collaborators derived to 
isolate the nonlocal and nonanalytic parts of scattering amplitudes that 
affect long-range potentials \cite{Donoghue:1994dn,Donoghue:1996mt}. 
These ``Donoghue Identities'' reduce internal massive propagators to 
delta functions, which causes higher-point contributions to 2-particle 
scattering to assume a form that can be regarded as corrections to the 
1PI 2-point function of the massless exchange particle. The new technique 
was implemented at one loop order for quantum gravitational corrections 
to a massless scalar on flat space background to derive a result which is
indpendent of the gauge parameters $\alpha$ and $\beta$ 
\cite{Miao:2017feh}. The same technique has just been applied to quantum
gravitational corrections to electrodynamics on flat space background
\cite{Katuwal:2021ljt}, and strenuous efforts are underway to generalize 
it to de Sitter background \cite{Glavan:2019msf}. The two flat space 
exercises demonstrate that gauge independent results typically show the 
same spacetime dependence as in a fixed gauge, but with different numerical 
coefficients. We expect that the same thing will be found on de Sitter 
background.

We should also comment on the breakdown of perturbation theory which 
must occur over the course of a very long period of inflation during which
$\ln(a)$ becomes larger than $1/\kappa H$. This would be relevant to
$\Lambda$-driven inflation in which inflation is driven by a positive
cosmological constant, without any scalar inflaton, and the self-gravitation
of inflationary gravitons gradually builds up to stop inflation
\cite{Tsamis:1996qq,Tsamis:2011ep}. Of course one cannot trust our
one loop result result (\ref{lateu1}) when $\kappa^2 H^2 \ln^2(a) \sim 1$ 
because unknown higher loop corrections could be equally or even more 
important. What is needed is a way to sum up the leading logarithms at each 
order. Starobinsky has developed a stochastic formalism that accomplishes 
this for the similar large logarithms produced in scalar potential models 
\cite{Starobinsky:1986fx,Starobinsky:1994bd}. This technique captures large 
logarithms that arise from the ``tail'' part of the propagator 
\cite{Tsamis:2005hd}, but it fails for large logarithms which have an 
ultraviolet origin, as can sometimes happen with the derivative
interactions of quantum gravity \cite{Miao:2008sp}. The renormalization
group was designed to sum up large logarithms from the ultraviolet, but
it fails to recover large logarithms from the ``tail'' part of the 
propagator \cite{Woodard:2008yt}. Our one loop result (\ref{lateu1})
is a purely ``tail'' effect, but it is not known what happens at higher 
loops, and one loop quantum gravitational corrections to electrodynamics 
are known to derive from both sources \cite{Miao:2018bol}. It would be 
interesting to try summing all large quantum gravitational logarithms by 
combining a variant of the renormlization group with some version of 
Starobinsky's technique.

\vspace{0cm}

\centerline{\bf Acknowledgements}

This work was partially supported by the European Union's Horizon 
2020 Programme under grant agreement 669288-SM-GRAV-ERC-2014-ADG;
by NSF grant 1912484; and by the UF's Institute for Fundamental Theory.


\begin{thebibliography}{99}

\bibitem{Starobinsky:1979ty}
A.~A.~Starobinsky,
JETP Lett. \textbf{30}, 682-685 (1979)

\bibitem{Starobinsky:1985ww}
A.~A.~Starobinsky,
Sov. Astron. Lett. \textbf{11}, 133 (1985)

\bibitem{Tsamis:1992xa}
N.~C.~Tsamis and R.~P.~Woodard,
Commun. Math. Phys. \textbf{162}, 217-248 (1994)
doi:10.1007/BF02102015

\bibitem{Woodard:2004ut}
R.~P.~Woodard,
[arXiv:gr-qc/0408002 [gr-qc]].

\bibitem{Iliopoulos:1998wq}
J.~Iliopoulos, T.~N.~Tomaras, N.~C.~Tsamis and R.~P.~Woodard,
Nucl. Phys. B \textbf{534}, 419-446 (1998)
doi:10.1016/S0550-3213(98)00528-8
[arXiv:gr-qc/9801028 [gr-qc]].

\bibitem{Abramo:2001dc}
L.~R.~Abramo and R.~P.~Woodard,
Phys. Rev. D \textbf{65}, 063515 (2002)
doi:10.1103/PhysRevD.65.063515
[arXiv:astro-ph/0109272 [astro-ph]].

\bibitem{Tsamis:1996qk}
N.~C.~Tsamis and R.~P.~Woodard,
Phys. Rev. D \textbf{54}, 2621-2639 (1996)
doi:10.1103/PhysRevD.54.2621
[arXiv:hep-ph/9602317 [hep-ph]].

\bibitem{Tan:2021ibs}
L.~Tan, N.~C.~Tsamis and R.~P.~Woodard,
[arXiv:2103.08547 [gr-qc]].

\bibitem{Tsamis:1996qm}
N.~C.~Tsamis and R.~P.~Woodard,
Annals Phys. \textbf{253}, 1-54 (1997)
doi:10.1006/aphy.1997.5613
[arXiv:hep-ph/9602316 [hep-ph]].

\bibitem{Onemli:2002hr}
V.~K.~Onemli and R.~P.~Woodard,
Class. Quant. Grav. \textbf{19}, 4607 (2002)
doi:10.1088/0264-9381/19/17/311
[arXiv:gr-qc/0204065 [gr-qc]].

\bibitem{Onemli:2004mb}
V.~K.~Onemli and R.~P.~Woodard,
Phys. Rev. D \textbf{70}, 107301 (2004)
doi:10.1103/PhysRevD.70.107301
[arXiv:gr-qc/0406098 [gr-qc]].

\bibitem{Schwinger:1960qe}
J.~S.~Schwinger,
J. Math. Phys. \textbf{2}, 407-432 (1961)
doi:10.1063/1.1703727

\bibitem{Mahanthappa:1962ex}
K.~T.~Mahanthappa,
Phys. Rev. \textbf{126}, 329-340 (1962)
doi:10.1103/PhysRev.126.329

\bibitem{Bakshi:1962dv}
P.~M.~Bakshi and K.~T.~Mahanthappa,
J. Math. Phys. \textbf{4}, 1-11 (1963)
doi:10.1063/1.1703883

\bibitem{Bakshi:1963bn}
P.~M.~Bakshi and K.~T.~Mahanthappa,
J. Math. Phys. \textbf{4}, 12-16 (1963)
doi:10.1063/1.1703879

\bibitem{Keldysh:1964ud}
L.~V.~Keldysh,
Zh. Eksp. Teor. Fiz. \textbf{47}, 1515-1527 (1964)

\bibitem{Chou:1984es}
K.~c.~Chou, Z.~b.~Su, B.~l.~Hao and L.~Yu,
Phys. Rept. \textbf{118}, 1-131 (1985)
doi:10.1016/0370-1573(85)90136-X

\bibitem{Jordan:1986ug}
R.~D.~Jordan,
Phys. Rev. D \textbf{33}, 444-454 (1986)
doi:10.1103/PhysRevD.33.444

\bibitem{Calzetta:1986ey}
E.~Calzetta and B.~L.~Hu,
Phys. Rev. D \textbf{35}, 495 (1987)
doi:10.1103/PhysRevD.35.495

\bibitem{Ford:2004wc}
L.~H.~Ford and R.~P.~Woodard,
Class. Quant. Grav. \textbf{22}, 1637-1647 (2005)
doi:10.1088/0264-9381/22/9/011
[arXiv:gr-qc/0411003 [gr-qc]].

\bibitem{Mora:2013ypa}
P.~J.~Mora, N.~C.~Tsamis and R.~P.~Woodard,
JCAP \textbf{10}, 018 (2013)
doi:10.1088/1475-7516/2013/10/018
[arXiv:1307.1422 [gr-qc]].

\bibitem{Prokopec:2002uw}
T.~Prokopec, O.~Tornkvist and R.~P.~Woodard,
Annals Phys. \textbf{303}, 251-274 (2003)
doi:10.1016/S0003-4916(03)00004-6
[arXiv:gr-qc/0205130 [gr-qc]].

\bibitem{Duffy:2005ue}
L.~D.~Duffy and R.~P.~Woodard,
Phys. Rev. D \textbf{72}, 024023 (2005)
doi:10.1103/PhysRevD.72.024023
[arXiv:hep-ph/0505156 [hep-ph]].

\bibitem{Tan:2022xpn}
L.~Tan, N.~C.~Tsamis and R.~P.~Woodard,
[arXiv:2206.11467 [gr-qc]].

\bibitem{Weinberg:2005vy}
S.~Weinberg,
Phys. Rev. D \textbf{72}, 043514 (2005)
doi:10.1103/PhysRevD.72.043514
[arXiv:hep-th/0506236 [hep-th]].

\bibitem{Weinberg:2006ac}
S.~Weinberg,
Phys. Rev. D \textbf{74}, 023508 (2006)
doi:10.1103/PhysRevD.74.023508
[arXiv:hep-th/0605244 [hep-th]].

\bibitem{Planck:2018vyg}
N.~Aghanim \textit{et al.} [Planck],
Astron. Astrophys. \textbf{641}, A6 (2020)
doi:10.1051/0004-6361/201833910
[arXiv:1807.06209 [astro-ph.CO]].

\bibitem{Loeb:2003ya}
A.~Loeb and M.~Zaldarriaga,
Phys. Rev. Lett. \textbf{92}, 211301 (2004)
doi:10.1103/PhysRevLett.92.211301
[arXiv:astro-ph/0312134 [astro-ph]].

\bibitem{Furlanetto:2006jb}
S.~Furlanetto, S.~P.~Oh and F.~Briggs,
Phys. Rept. \textbf{433}, 181-301 (2006)
doi:10.1016/j.physrep.2006.08.002
[arXiv:astro-ph/0608032 [astro-ph]].

\bibitem{Masui:2010cz}
K.~W.~Masui and U.~L.~Pen,
Phys. Rev. Lett. \textbf{105}, 161302 (2010)
doi:10.1103/PhysRevLett.105.161302
[arXiv:1006.4181 [astro-ph.CO]].

\bibitem{Capper:1979ej}
D.~M.~Capper,
J. Phys. A \textbf{13}, 199 (1980)
doi:10.1088/0305-4470/13/1/022

\bibitem{Glavan:2015ura}
D.~Glavan, S.~P.~Miao, T.~Prokopec and R.~P.~Woodard,
Class. Quant. Grav. \textbf{32}, no.19, 195014 (2015)
doi:10.1088/0264-9381/32/19/195014
[arXiv:1504.00894 [gr-qc]].

\bibitem{Tsamis:2005je}
N.~C.~Tsamis and R.~P.~Woodard,
Annals Phys. \textbf{321}, 875-893 (2006)
doi:10.1016/j.aop.2005.08.004
[arXiv:gr-qc/0506056 [gr-qc]].

\bibitem{Miao:2005am}
S.~P.~Miao and R.~P.~Woodard,
Class. Quant. Grav. \textbf{23}, 1721-1762 (2006)
doi:10.1088/0264-9381/23/5/016
[arXiv:gr-qc/0511140 [gr-qc]].

\bibitem{Kahya:2007bc}
E.~O.~Kahya and R.~P.~Woodard,
Phys. Rev. D \textbf{76}, 124005 (2007)
doi:10.1103/PhysRevD.76.124005
[arXiv:0709.0536 [gr-qc]].

\bibitem{Miao:2012bj}
S.~P.~Miao,
Phys. Rev. D \textbf{86}, 104051 (2012)
doi:10.1103/PhysRevD.86.104051
[arXiv:1207.5241 [gr-qc]].

\bibitem{Leonard:2013xsa}
K.~E.~Leonard and R.~P.~Woodard,
Class. Quant. Grav. \textbf{31}, 015010 (2014)
doi:10.1088/0264-9381/31/1/015010
[arXiv:1304.7265 [gr-qc]].

\bibitem{Boran:2014xpa}
S.~Boran, E.~O.~Kahya and S.~Park,
Phys. Rev. D \textbf{90}, no.12, 124054 (2014)
doi:10.1103/PhysRevD.90.124054
[arXiv:1409.7753 [gr-qc]].

\bibitem{Boran:2017fsx}
S.~Boran, E.~O.~Kahya and S.~Park,
Phys. Rev. D \textbf{96}, no.2, 025001 (2017)
doi:10.1103/PhysRevD.96.025001
[arXiv:1704.05880 [gr-qc]].

\bibitem{Glavan:2020gal}
D.~Glavan, S.~P.~Miao, T.~Prokopec and R.~P.~Woodard,
Phys. Rev. D \textbf{101}, no.10, 106016 (2020)
doi:10.1103/PhysRevD.101.106016
[arXiv:2003.02549 [gr-qc]].

\bibitem{Mora:2012zi}
P.~J.~Mora, N.~C.~Tsamis and R.~P.~Woodard,
J. Math. Phys. \textbf{53}, 122502 (2012)
doi:10.1063/1.4764882
[arXiv:1205.4468 [gr-qc]].

\bibitem{Glavan:2016bvp}
D.~Glavan, S.~P.~Miao, T.~Prokopec and R.~P.~Woodard,
Class. Quant. Grav. \textbf{34}, no.8, 085002 (2017)
doi:10.1088/1361-6382/aa61da
[arXiv:1609.00386 [gr-qc]].

\bibitem{Wang:2014tza}
C.~L.~Wang and R.~P.~Woodard,
Phys. Rev. D \textbf{91}, no.12, 124054 (2015)
doi:10.1103/PhysRevD.91.124054
[arXiv:1408.1448 [gr-qc]].

\bibitem{Donoghue:1993eb}
J.~F.~Donoghue,
Phys. Rev. Lett. \textbf{72}, 2996-2999 (1994)
doi:10.1103/PhysRevLett.72.2996
[arXiv:gr-qc/9310024 [gr-qc]].

\bibitem{Donoghue:1994dn}
J.~F.~Donoghue,
Phys. Rev. D \textbf{50}, 3874-3888 (1994)
doi:10.1103/PhysRevD.50.3874
[arXiv:gr-qc/9405057 [gr-qc]].

\bibitem{Bjerrum-Bohr:2002fji}
N.~E.~J.~Bjerrum-Bohr, J.~F.~Donoghue and B.~R.~Holstein,
Phys. Rev. D \textbf{68}, 084005 (2003)
[erratum: Phys. Rev. D \textbf{71}, 069904 (2005)]
doi:10.1103/PhysRevD.68.084005
[arXiv:hep-th/0211071 [hep-th]].

\bibitem{Bjerrum-Bohr:2002gqz}
N.~E.~J.~Bjerrum-Bohr, J.~F.~Donoghue and B.~R.~Holstein,
Phys. Rev. D \textbf{67}, 084033 (2003)
[erratum: Phys. Rev. D \textbf{71}, 069903 (2005)]
doi:10.1103/PhysRevD.71.069903
[arXiv:hep-th/0211072 [hep-th]].

\bibitem{Bjerrum-Bohr:2002aqa}
N.~E.~J.~Bjerrum-Bohr,
Phys. Rev. D \textbf{66}, 084023 (2002)
doi:10.1103/PhysRevD.66.084023
[arXiv:hep-th/0206236 [hep-th]].

\bibitem{Miao:2017feh}
S.~P.~Miao, T.~Prokopec and R.~P.~Woodard,
Phys. Rev. D \textbf{96}, no.10, 104029 (2017)
doi:10.1103/PhysRevD.96.104029
[arXiv:1708.06239 [gr-qc]].

\bibitem{Donoghue:1996mt}
J.~F.~Donoghue and T.~Torma,
Phys. Rev. D \textbf{54}, 4963-4972 (1996)
doi:10.1103/PhysRevD.54.4963
[arXiv:hep-th/9602121 [hep-th]].

\bibitem{Katuwal:2021ljt}
S.~Katuwal and R.~P.~Woodard,
[arXiv:2107.13341 [gr-qc]].

\bibitem{Glavan:2019msf}
D.~Glavan, S.~P.~Miao, T.~Prokopec and R.~P.~Woodard,
JHEP \textbf{10}, 096 (2019)
doi:10.1007/JHEP10(2019)096
[arXiv:1908.06064 [gr-qc]].

\bibitem{Tsamis:1996qq}
N.~C.~Tsamis and R.~P.~Woodard,
Nucl. Phys. B \textbf{474}, 235-248 (1996)
doi:10.1016/0550-3213(96)00246-5
[arXiv:hep-ph/9602315 [hep-ph]].

\bibitem{Tsamis:2011ep}
N.~C.~Tsamis and R.~P.~Woodard,
Int. J. Mod. Phys. D \textbf{20}, 2847-2851 (2011)
doi:10.1142/S0218271811020652
[arXiv:1103.5134 [gr-qc]].

\bibitem{Starobinsky:1986fx}
A.~A.~Starobinsky,
Lect. Notes Phys. \textbf{246}, 107-126 (1986)
doi:10.1007/3-540-16452-9\_6

\bibitem{Starobinsky:1994bd}
A.~A.~Starobinsky and J.~Yokoyama,
Phys. Rev. D \textbf{50}, 6357-6368 (1994)
doi:10.1103/PhysRevD.50.6357
[arXiv:astro-ph/9407016 [astro-ph]].

\bibitem{Tsamis:2005hd}
N.~C.~Tsamis and R.~P.~Woodard,
Nucl. Phys. B \textbf{724}, 295-328 (2005)
doi:10.1016/j.nuclphysb.2005.06.031
[arXiv:gr-qc/0505115 [gr-qc]].

\bibitem{Miao:2008sp}
S.~P.~Miao and R.~P.~Woodard,
Class. Quant. Grav. \textbf{25}, 145009 (2008)
doi:10.1088/0264-9381/25/14/145009
[arXiv:0803.2377 [gr-qc]].

\bibitem{Woodard:2008yt}
R.~P.~Woodard,
Phys. Rev. Lett. \textbf{101}, 081301 (2008)
doi:10.1103/PhysRevLett.101.081301
[arXiv:0805.3089 [gr-qc]].

\bibitem{Miao:2018bol}
S.~P.~Miao, T.~Prokopec and R.~P.~Woodard,
Phys. Rev. D \textbf{98}, no.2, 025022 (2018)
doi:10.1103/PhysRevD.98.025022
[arXiv:1806.00742 [gr-qc]].

\end{thebibliography}
\end{document}